\documentclass{article}
\usepackage{ijcai25}

\usepackage{times}
\usepackage{soul}
\usepackage{url}
\usepackage[hidelinks]{hyperref}
\usepackage[utf8]{inputenc}
\usepackage[small]{caption}
\usepackage{graphicx}
\usepackage{amsmath}
\usepackage{amsthm}
\usepackage{booktabs}
\usepackage{algorithm}
\usepackage{algorithmic}
\usepackage[switch]{lineno}

\usepackage{bm}
\usepackage{amsmath}
\usepackage{amssymb}
\usepackage{amsthm}
\usepackage{subfig}
\usepackage{graphicx}
\usepackage{caption}
\usepackage{booktabs}
\catcode`\@ = 11
\newdimen\@InsertBoxMargin
\newcount\@numlines    
\newcount\@linesleft   
\def\ParShape{%
    \@numlines = 0
    \def\@parshapedata{ }
    \afterassignment\@beginParShape
    \@linesleft
}%
\def\@beginParShape{%
    \ifnum \@linesleft = 0
      \let\@whatnext = \@endParShape
    \else
      \let\@whatnext = \@readnextline
    \fi
    \@whatnext
}%
\def\@endParShape{%
    \global\parshape = \@numlines \@parshapedata
}%
\def\@readnextline#1 #2 #3 {
    \ifnum #1 > 0
      \bgroup  
        \dimen0 = \hsize
        \advance \dimen0 by -#2  
        \advance \dimen0 by -#3  
        \count0 = 0
        \loop
          \global\edef\@parshapedata{%
            \@parshapedata    
            #2                
            \space            
            \the\dimen0       
            \space            
          }%
          \advance \count0 by 1
          \ifnum \count0 < #1
        \repeat
      \egroup
      \advance \@numlines by #1
    \fi
    \advance \@linesleft by -1
    \@beginParShape
}%
\newbox\@boxcontent     
\newcount\@numnormal    
\newdimen\@framewidth   
\newdimen\@wherebottom  
\newif\if@byframe       
\@byframefalse
\def\InsertBoxC#1{%
  \leavevmode
  \vadjust{
    \vskip \@InsertBoxMargin
    \hbox to \hsize{\hss#1\hss}
    \vskip \@InsertBoxMargin
  }%
}%
\def\InsertBoxL#1#2{%
  \@numnormal = #1
  \setbox\@boxcontent = \hbox{#2}%
  \let\@side = 0
  \futurelet \@optionalparameter \@InsertBox
}
\def\InsertBoxR#1#2{%
  \@numnormal = #1
  \setbox\@boxcontent = \hbox{#2}%
  \let\@side = 1
  \futurelet \@optionalparameter \@InsertBox
}%
\def\@InsertBox{%
  \ifx \@optionalparameter [
    \let\@whatnext = \@@InsertBoxCorrection
  \else
    \let\@whatnext = \@@InsertBoxNoCorrection
  \fi
  \@whatnext
}%
\def\@@InsertBoxCorrection[#1]{%
  \ifx \@side 0
    \@@InsertBox{#1}{0}{{\the\@framewidth} 0cm}%
  \else
    \@@InsertBox{#1}{1}{0cm {\the\@framewidth}}%
  \fi
}%
\def\@@InsertBoxNoCorrection{%
  \@@InsertBoxCorrection[0]%
}%
\def\@@InsertBox#1#2#3{%
  \MoveBelowBox
  \@byframetrue
  \@wherebottom = \baselineskip
  \multiply \@wherebottom by \@numnormal
  \advance \@wherebottom by 2\@InsertBoxMargin
  \advance \@wherebottom by \ht\@boxcontent
  \advance \@wherebottom by \pagetotal
  \ifdim \pagetotal = 0cm
    \advance \@wherebottom by -\baselineskip  
  \fi
  \advance \@wherebottom by #1\baselineskip
  \@framewidth = \wd\@boxcontent
  \advance \@framewidth by \@InsertBoxMargin
  \bgroup  
    \ifdim \pagetotal = 0cm
      \dimen0 = \vsize
    \else
      \dimen0 = \pagegoal
    \fi
    \ifdim \@wherebottom > \dimen0
      \immediate\write16{+--------------------------------------------------------------+}%
      \immediate\write16{| The box will not fit in the page. Please, re-edit your text. |}%
      \immediate\write16{+--------------------------------------------------------------+}%
      \vrule width \overfullrule
    \fi
  \egroup
  \prevgraf = 0
  \vbox to 0cm{%
    \dimen0 = \baselineskip
    \multiply \dimen0 by \@numnormal
    \advance \dimen0 by -\baselineskip
    \setbox0 = \hbox{y}%
    \vskip \dp0
    \vskip \dimen0
    \vskip \@InsertBoxMargin
    \ifnum #2 = 1
      \vtop{\noindent \hbox to \hsize{\hss \box\@boxcontent}}%
    \else
      \vtop{\noindent \box\@boxcontent}%
    \fi
    \vss
  }%
  \vglue -\parskip
  \vskip -\baselineskip
  \everypar = {%
    \ifdim \pagetotal < \@wherebottom
      \bgroup  
        \dimen0 = \@wherebottom
        \advance \dimen0 by -\pagetotal
        \divide \dimen0 by \baselineskip
        \count1 = \dimen0
        \advance \count1 by 1
        \advance \count1 by -\@numnormal
        \ifnum #2 = 1
          \ParShape = 3
                      {\the\@numnormal}   0cm   0cm
                      {\the\count1}       0cm   {\the\@framewidth}
                      1                   0cm   0cm
        \else
          \ParShape = 3
                      {\the\@numnormal}   0cm                  0cm
                      {\the\count1}       {\the\@framewidth}   0cm
                      1                   0cm                  0cm
        \fi
      \egroup
    \else
      \@restore@    
    \fi
  }%
  \def\par{%
      \endgraf
      \global\advance \@numnormal by -\prevgraf
      \ifnum \@numnormal < 0
        \global\@numnormal = 0
      \fi
      \prevgraf = 0
  }%
}%
\def\MoveBelowBox{%
  \par
  \if@byframe
    \global\advance \@wherebottom by -\pagetotal
    \ifdim \@wherebottom > 0cm
      \vskip \@wherebottom
    \fi
    \@restore@
  \fi
}%
\def\@restore@{%
    \global\@wherebottom = 0cm
    \global\@byframefalse
    \global\everypar = {}%
    \global\let \par = \endgraf
    \global\parshape = 1 0cm \hsize
}%
\ifx \documentclass \@Dont@Know@What@It@Is@
\else
  \let \pageno = \c@page
\fi

\catcode`\@ = 12

\newtheorem{definition}{Definition}
\newtheorem{theorem}{Theorem}

\newtheorem{appendix_lemma}{Lemma}
\newtheorem{appendix_theorem}{Theorem}

\title{Asynchronous Credit Assignment for Multi-Agent Reinforcement Learning
	\thanks{Accepted at IJCAI-25. Copyright \textcopyright~IJCAI.}
}

\author{
	Yongheng Liang$^{1,2}$
	\and
	Hejun Wu$^{1,2}$\thanks{Corresponding author}\and
	Haitao Wang$^{1,2}$\And
	Hao Cai$^3$\\
	\affiliations
	$^1$School of Computer Science and Engineering, Sun Yat-sen University, Guangzhou, China\\
	$^2$Guangdong Key Laboratory of Big Data Analysis and Processing, Guangzhou, Guangdong, China\\
	$^3$College of Mathematics and Computer Science, Shantou University, Shantou, China\\
	\emails
	\{liangyh38, wanght39\}@mail2.sysu.edu.cn,
	wuhejun@mail.sysu.edu.cn,
	haocai@stu.edu.cn
}

\begin{document}
	
	\maketitle
	
	\begin{abstract}
		Credit assignment is a critical problem in multi-agent reinforcement learning (MARL), aiming to identify agents' marginal contributions for optimizing cooperative policies. Current credit assignment methods typically assume synchronous decision-making among agents. However, many real-world scenarios require agents to act asynchronously without waiting for others. This asynchrony introduces conditional dependencies between actions, which pose great challenges to current methods. To address this issue, we propose an asynchronous credit assignment framework, incorporating a Virtual Synchrony Proxy (VSP) mechanism and a Multiplicative Value Decomposition (MVD) algorithm. VSP enables physically asynchronous actions to be virtually synchronized during credit assignment. We theoretically prove that VSP preserves both task equilibrium and algorithm convergence. Furthermore, MVD leverages multiplicative interactions to effectively model dependencies among asynchronous actions, offering theoretical advantages in handling asynchronous tasks. Extensive experiments show that our framework consistently outperforms state-of-the-art MARL methods on challenging tasks while providing improved interpretability for asynchronous cooperation.

	\end{abstract}
	
	\section{Introduction}

Multi-agent reinforcement learning (MARL) is promising for many cooperative tasks, such as video games \cite{marl1} and collaborative control \cite{marl4}. MARL typically assumes a synchronous decision-making setting, where all agents make decisions simultaneously and their joint actions have the same duration. This assumption simplifies the overall learning process \cite{decpomdp}.

\begin{figure}[!t]
	\centering
	\subfloat[Waiting]{
		\includegraphics[width=0.22\textwidth]{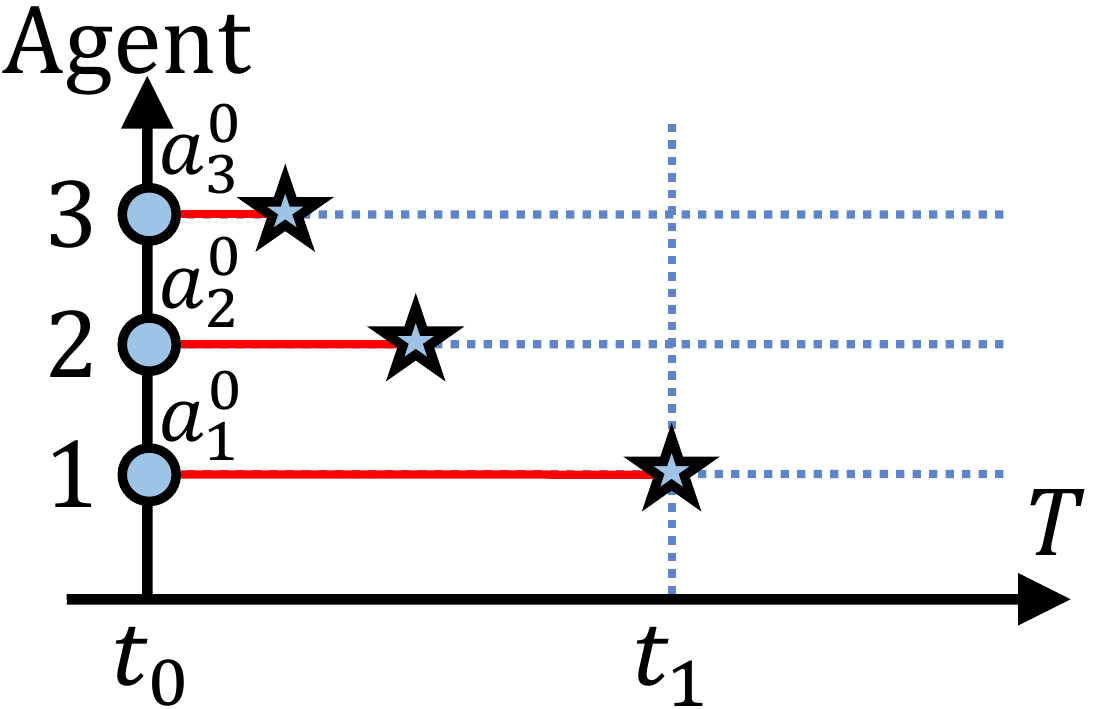}
		\label{fig:intro5}
	}
	\subfloat[Discarding]{
		\includegraphics[width=0.22\textwidth]{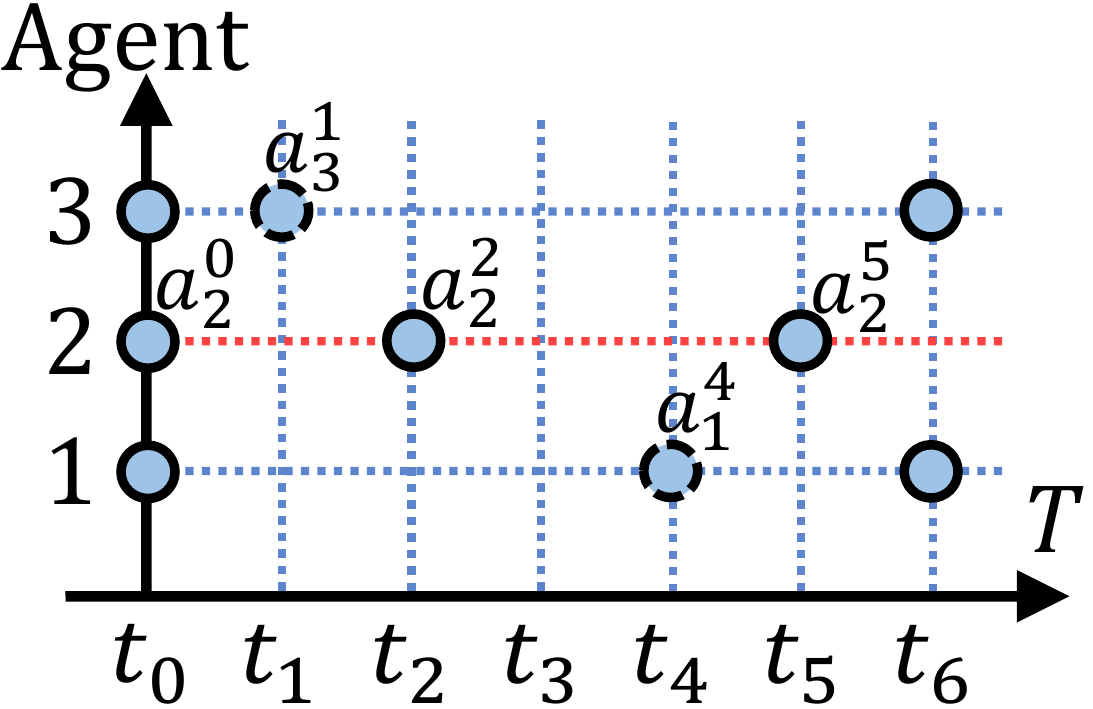}
		\label{fig:intro1}
	}
	\vfil
	\subfloat[Padding]{
		\includegraphics[width=0.22\textwidth]{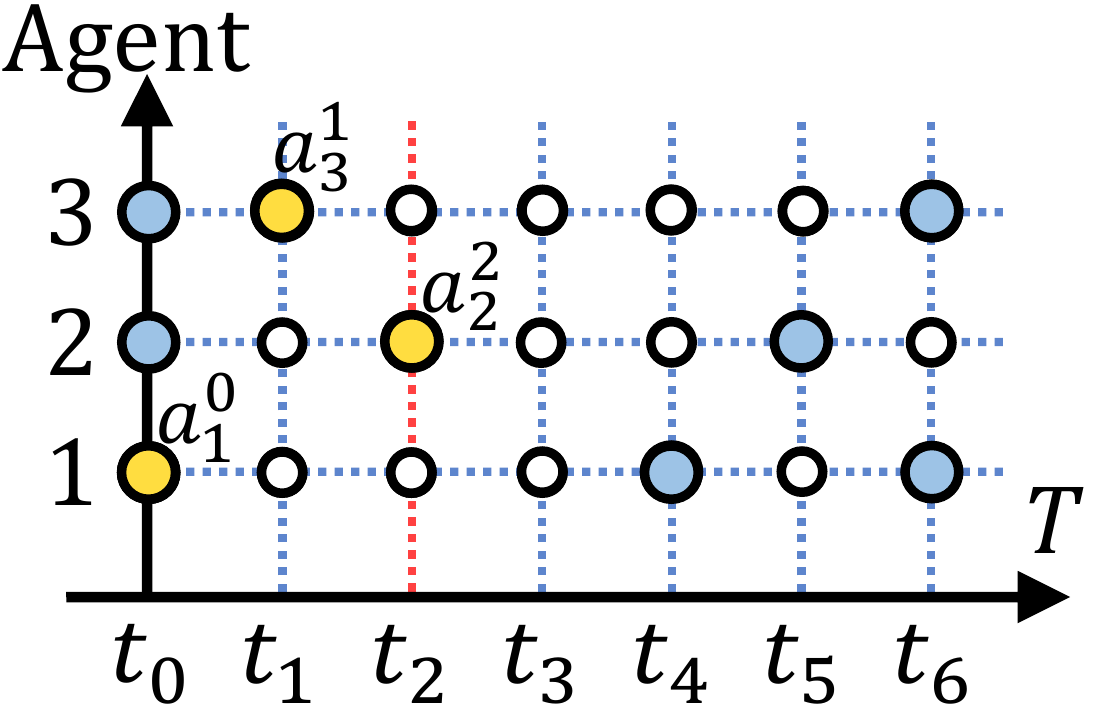}
		\label{fig:intro3}
	}
	\subfloat[Our framework]{
		\includegraphics[width=0.22\textwidth]{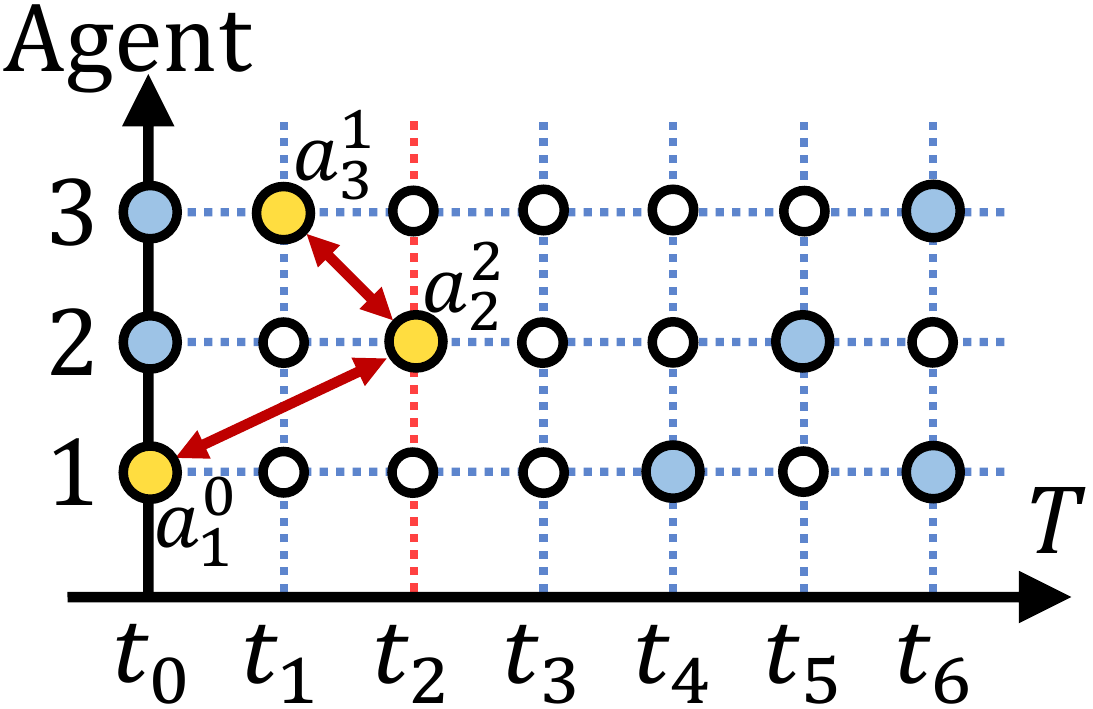}
		\label{fig:intro4}
	}
	\caption{Illustration of various asynchronous MARL frameworks. Blue/Yellow circles denote agent $\#x$'s action $a_x^y$ at time step $t_y$. Small white circles denote padding actions. Stars denote action completion. (a) Agents $\#2$ and $\#3$ must wait for agent $\#1$ to finish at $t_1$ before making next decisions. (b) Agent $\#2$ disregards action $a_1^4$ from $t_2$ to $t_5$. (c) Credit at $t_2$ is attributed to the padding actions. (d) Our proposed framework captures the interactions among asynchronous decisions being executed at $t_2$.}
	\label{intro}
\end{figure}

Despite the success of MARL in synchronous settings, real-world tasks often exhibit asynchrony, i.e., agents cannot complete their atomic actions simultaneously, because of hardware constraints or the nature of agents' actions \cite{caac,asmhppo}, as shown in Figure \ref{fig:intro5}. To address this issue, researchers have proposed two primary mechanisms as follows. $Discarding$: Agents with varying time step lengths collect data and update policies only when they make decisions \cite{maciaicc}, as shown in Figure \ref{fig:intro1}. $Padding$: Define a time step as the smallest indivisible time unit and use padding actions to transform asynchronous tasks into synchronous ones, so as to apply existing MARL \cite{varlenmarl}, as shown in Figure \ref{fig:intro3}.

Nevertheless, both discarding and padding struggle to address complex asynchronous cooperative tasks. Their failure is due to an inability to resolve the credit assignment problem, caused by the following two limitations. (1) \textbf{Bias introduced in evaluating the global impacts of asynchronous actions}. On one hand, the discarded information leads to biased evaluations of the impacts of decisions from other agents. As shown in Figure \ref{fig:intro1}, agent $\#2$ discards the information of action $a_3^1$ and $a_1^4$, neglecting their impacts on agent $\#2$'s transition from $t_2$ to $t_5$. On the other hand, the padding actions introduce bias in credit assignment. As shown in Figure \ref{fig:intro3}, the algorithm assigns credit at $t_2$ mainly to the two padding actions, rather than actions $a_1^0$ and $a_3^1$. (2) \textbf{Inability to model conditional dependencies between asynchronous actions}. In MARL, value decomposition (VD) and its variants \cite{vdn,qmix} are widely used synchronous credit assignment methods. They learn the marginal contribution of each agent and decompose the global Q-value $Q_{tot}$ into individual agent-wise utilities $Q_i$ to guide agents’ behaviors. However, most VD algorithms have limitations in accounting for higher-order interactions \cite{na2q}. They fail to capture the conditional dependencies between decisions made by agents asynchronously, such as the dependency of the current decision $a_2^2$ on the actions being executed, $a_1^0$ and $a_3^1$, as shown in Figure \ref{fig:intro3}.

In this paper, we propose an asynchronous credit assignment framework that incorporates a Virtual Synchrony Proxy (VSP) mechanism and a Multiplicative Value Decomposition (MVD) algorithm. Inspired by virtual synchrony in distributed systems \cite{vs}, our VSP introduces virtual proxies to migrate asynchronous actions to a unified time step. This allows actions $a_1^0$, $a_2^2$, and $a_3^1$ in Figure \ref{fig:intro4} to appear synchronized at $t_2$, facilitating better capturing their global impacts. We have proven that VSP preserves both the task equilibrium and the algorithm convergence. Based on VSP, we derive the multiplicative value decomposition formula as well as its higher-order forms and propose MVD. Our MVD leverages multiplicative interactions \cite{miold,mi} to capture dependencies among asynchronous actions and we demonstrate its superior representational capacity compared to traditional VD algorithms. Moreover, we present three practical implementations of MVD. We evaluate MVD on a modified asynchronous variant of the classic MARL benchmark SMAC \cite{smac}, along with two prominent asynchronous benchmarks: Overcooked \cite{overcooked} and POAC \cite{poac}. Extensive experimental results show that MVD achieves considerable performance improvements in complex scenarios and provides easy-to-understand interaction processes among asynchronous decisions.

Our contributions are summarized as follows: (1) We propose an asynchronous credit assignment framework with VSP and MVD, capable of capturing high-order dependencies among asynchronous actions. (2) Theoretically, we prove the correctness of VSP and demonstrate the advantages of MVD. (3) Experimentally, we show MVD’s effectiveness across three asynchronous tasks, achieving significant performance gains and interpretable interaction processes.

	\section{Related Works}

Despite the significant progress in MARL, most existing works rely on the premise of synchronous decision-making which does not reflect reality in many practical applications. The easiest way to adapt MARL from synchronous to asynchronous decision-making is to split actions into sub-actions or wait for others to finish before making the next decisions. Evidently, these methods raise training costs and lower efficiency. Thus, several works have been conducted to exploit the strengths of MARL in asynchronous settings.

The discarding type methods recognize asynchronous actions with varying durations as a whole and focus solely on the decision information. ASM-PPO \cite{asmppo} and ASM-HPPO \cite{asmhppo} propose that each agent collects its own decision information and utilize MAPPO \cite{mappo} for training. MAC IAICC \cite{maciaicc} treats asynchronous actions as macro-actions \cite{macro1} and models the task as a MacDec-POMDP \cite{macdec2}. CAAC \cite{caac} focuses on the bus holding control \cite{budholding} and utilizes a graph attention network to capture the influence of agents' asynchronous decisions.

The padding type methods transform asynchronous problems into synchronous ones through padding action, thereby obtaining Dec-POMDP \cite{decpomdp} and applying existing MARL methods. Since Dec-POMDP requires the collection of decision information from all agents at each time step, the padding action can be used as a substitute for the decision information of agents that are executing actions. VarLenMARL \cite{varlenmarl} employs the most recent action for padding during the collection of joint transitions. EXP-Ms \cite{fever} considers ongoing actions as idle, treating them as blank actions. 

However, there still remains a lack of theoretical and visual analysis on asynchronous credit assignments, hindering the resolution of complex asynchronous cooperative tasks.

\section{Preliminaries}

\subsection{Dec-POMDP}

A fully cooperative multi-agent task with synchronous decision-making is typically formulated as a Dec-POMDP. Dec-POMDP is defined as a tuple $\langle \mathcal{N, S, A, P}, r, O, \Omega, \gamma \rangle$, where $\mathcal{N}$ is a set of $n$ agents and $s \in \mathcal{S}$ is a global state of the environment. At each time step, each agent $i \in \mathcal{N}$ obtains its own observation $o_i \in \Omega$ determined by the partial observation $O(s, i)$ and selects an action $a_i \in \mathcal{A}$ to form a joint action $\bm{a} = [a_i]^n_{i=1} \in \mathcal{A}^n$. Subsequently, all agents simultaneously complete their actions, leading to the next state $s'$ through the transition function $\mathcal{P}(s'|s, \bm{a}): \mathcal{S} \times \mathcal{A}^n \rightarrow \mathcal{S}$ and to the global reward $r(s, \bm{a}): \mathcal{S} \times \mathcal{A}^n \rightarrow \mathbb{R}$. Each agent $i$ has its own policy $\pi_i(a_i|\tau_i): \mathcal{T} \times \mathcal{A} \rightarrow [0, 1]$ based on local action-observation history $\tau_i \in \mathcal{T}$. The objective of all agents is to find an optimal joint policy $\bm{\pi}=[\pi_i]^n_{i=1}$ and maximize the global value function $Q^{\bm{\pi}} = \mathbb{E}[\sum^\infty_{t=0} \gamma^t r^{t+1}]$ with a discount factor $\gamma \in [0, 1)$.

\subsection{Credit Assignment and Value Decomposition}

Credit assignment is a key challenge in designing reliable MARL methods \cite{review}. It focuses on attributing team success to individual agents based on their respective marginal contributions, aiming at collective policy optimization. VD algorithms are the most popular branches in MARL. They leverage global information to learn agents' contributions and decompose the global Q-value function $Q_{tot}(s,\bm{a})$ into individual utility functions $Q_i(\tau_i, a_i)$. In the execution phase, agents cooperate via their corresponding $Q_i(\tau_i, a_i)$, thereby realizing centralized training and decentralized execution (CTDE) \cite{ctde}. Traditional VD algorithms, including VDN \cite{vdn}, QMIX \cite{qmix}, and Qatten \cite{qatten}, can be represented by the following general additive interaction formulation \cite{dvd}:
\begin{equation}
	\label{eq:ai}
	Q_{tot} = Q_{tot}(s, Q_1, Q_2, \cdots, Q_n) = k_0 + \sum^n_{i=1}k_iQ_i,
\end{equation}
where $k_0$ is a constant and $k_i$ denotes the credit that reflects the contributions of $Q_i$ to $Q_{tot}$.

To capture high-order interactions that traditional VD algorithms ignored, NA$^2$Q \cite{na2q} employed a generalized additive model (GAM) \cite{gam}:
\begin{equation}
	\begin{aligned}
		Q_{tot} = f_0 + \sum^n_{i=1}k_if_i^1(Q_i) + \cdots + \sum_{j \in \mathcal{D}_l}k_jf_j^l(Q_j) \\
		+ \cdots + k_{1 \cdots n} f_{1 \cdots n}^n(Q_1,\cdots,Q_n),
	\end{aligned}
	\label{eq:na2q}
\end{equation}
where $f_0$ is a constant, $f_j^l$ captures $l$-order interactions among agents $j$ in $\mathcal{D}_l$. $\mathcal{D}_l$ is the set of all size-$l$ subsets of $\{1, \cdots, n\}$, $1 \le l \le n$. The utility $Q_j$ of agent $j$ is the input of $f_j^l$.

In order to maintain the consistency between local and global optimal actions after decomposition, these VD algorithms should satisfy the following individual-global-max (IGM) principle \cite{qtran}:
\begin{equation}
	\label{eq:igm}
	\arg\max_{\bm{a}}Q_{tot}(s,\bm{a}) =
	\left(
	\begin{matrix}
		\arg\max_{a_1}Q_1(\tau_1, a_1) \\
		\vdots \\
		\arg\max_{a_n}Q_n(\tau_n, a_n)
	\end{matrix}
	\right).
\end{equation}

For example, QMIX holds the monotonicity $\frac{\partial Q_{tot}}{\partial Q_i} \ge 0$ and achieves IGM between $Q_{tot}$ and $Q_i$. More details of VD and other credit assignment methods are in Appendix A.

	\section{Virtual Synchrony Proxy}

In asynchronous scenarios, system dynamics depend on actions taken at different times, yet existing methods fail to capture the global impacts of these asynchronous actions. The discarding mechanism overlooks the contributions of other agents' decisions. One padding approach uses the most recent action to synchronize decisions. For example, in Figure \ref{fig:intro3}, agent $\#1$ selects $a_1^0$ as its padding actions at $t_2$. However, this introduces ambiguity since it cannot distinguish whether $a_1^0$ is being continued or restarted at $t_2$. Another padding approach uses blank actions to distinguish between decision-making and action execution, thereby avoiding ambiguity. Nevertheless, as shown in Figure \ref{fig:intro3}, the credit for agent $\#1$ at $t_2$ is incorrectly assigned to the blank action rather than to $a_1^0$.

\begin{figure}[t]
	\centering
	\subfloat[Re-execution of Actions]{
		\includegraphics[width=0.21\textwidth]{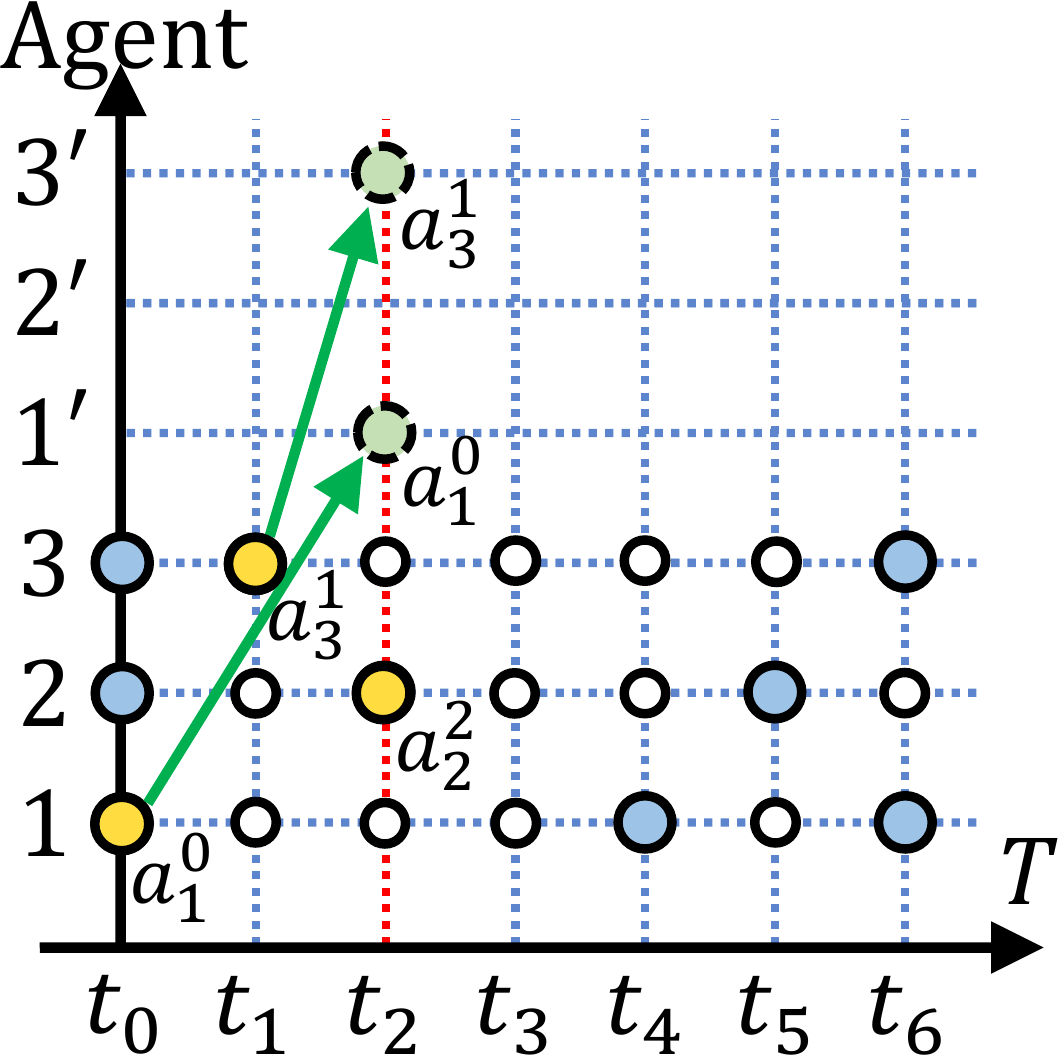}
		\label{fig:adex_1}
	}
	\subfloat[Policy Update via Proxy]{
		\includegraphics[width=0.21\textwidth]{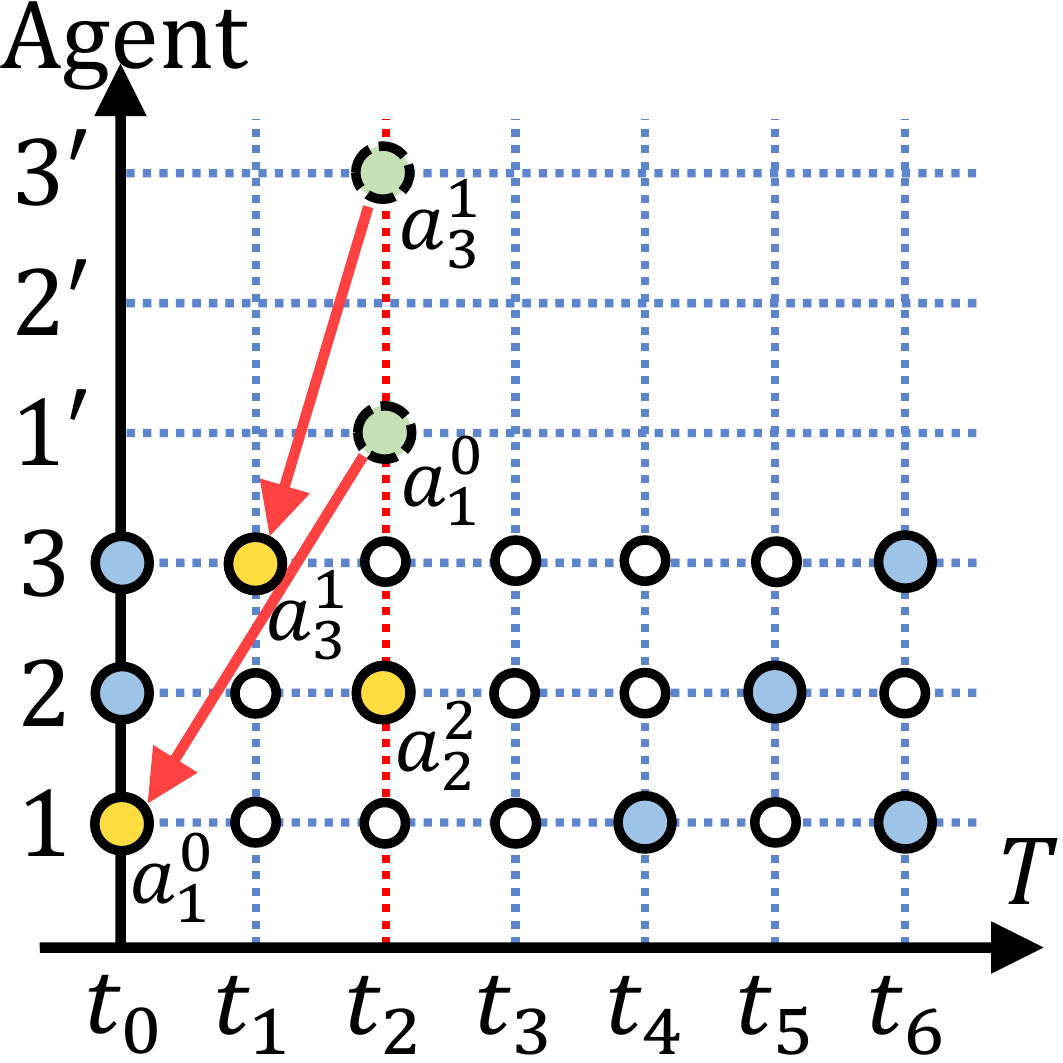}
		\label{fig:adex_2}
	}
	\caption{At $t_2$, $a_1^0$ and $a_3^1$ are executing. (a) Virtual proxies $\#1'$ and $\#3'$ are introduced to re-execute $a_1^0$ and $a_3^1$. (b) The policies of agent $\#1$ and $\#3$ for $a_1^0$ and $a_3^1$ are updated through their proxies.}
	\label{fig:adex}
\end{figure}

Our VSP is inspired by virtual synchrony, which offers an abstraction layer that ensures asynchronous messages are processed synchronously in distributed systems. Similarly, VSP employs virtual proxies to align asynchronous actions, allowing credit to be assigned at a unified time step. The general idea is as follows. To avoid ambiguity, VSP uses a \textbf{blank} action as padding when agent $i$ is continuing action $a_i$. Meanwhile, to ensure semantic consistency, VSP introduces a virtual proxy $i'$ at each time step when agent $i$ is executing action $a_i$. The proxy $i'$ shares the \textbf{same} policy as agent $i$ and re-executes action $a_i$. Therefore, when the policies of the proxies $\#1'$ and $\#3'$ are updated with assigned credit at $t_2$, the policies of the corresponding agents $\#1$ and $\#3$ are updated as well, as shown in Figure \ref{fig:adex}.

We integrate the VSP mechanism into Dec-POMDP to model asynchronous decision-making problems. Before presenting the detailed formulation, the frequently used definitions are outlined below. Asynchronous action refers to an action that requires multiple time steps to complete, e.g., action $a^0_1$ spans over four time steps in Figure \ref{fig:adex}. The observation for making a decision is denoted as $\tilde{o}_i$, referring to the most recent observation upon which an agent $i$ makes the decision to execute a new asynchronous action. $\tilde{a}_i$ denotes the most recent new decision made by the agent $i$.

\begin{definition}[Dec-POMDP with VSP]
	\label{def:adexpomdp}
	Dec-POMDP with VSP is a tuple $\langle \mathcal{\widehat{N}, I, \widehat{S}, A, \widehat{P}}, \widehat{r}, \widehat{O}, \mathcal{O_P}, \Omega, \gamma \rangle$, where $\mathcal{\widehat{N}} = \{ \mathcal{N}, \mathcal{N}' \}$, with $\mathcal{N}$ as the real agent set and $\mathcal{N}'$ as the virtual proxy set. At each time step $t$, the agents are divided into two disjoint sets: $i_d \in \mathcal{N}_d$, the agent making the decision, and $i_c \in \mathcal{N}_c$, the agent executing the action. Time step index $t$ is omitted for simplicity. Given an original state $s \in \mathcal{S}$, function $\mathcal{I}(s)$ returns a subset $\mathcal{N}'_c \subseteq \mathcal{N}'$, where proxy $i'_c \in \mathcal{N}'_c$ and agent $i_c \in \mathcal{N}_c$ form an real-virtual pair $\langle i_c, i'_c \rangle$. The agent $i_c$ and the proxy $i'_c$ in this pair have the same policy. $\widehat{s} = [s ; \bm{\tilde{o}}_c] \in \mathcal{\widehat{S}}$, where $[\cdot ; \cdot]$ is the concatenation operation. $s$ is the original state and $\bm{\tilde{o}}_c$ is obtained according to $\mathcal{O_P} (\bm{\tilde{o}}_c | s)$, which is the joint observation of all agents $i_c$ for making decisions. Each proxy $i'_c$ receives $o_{i'_c} = \tilde{o}_{i_c} \in \Omega$ according to $\widehat{O}(\widehat{s},i'_c)$ and selects action $a_{i'_c} = \tilde{a}_{i_c}$ to form $\widehat{\bm{a}} = [\bm{a}; \bm{\tilde{a}}_c] \in \mathcal{A}^{n+n'_c}$, where $n'_c = |\mathcal{N}'_c|$, $\bm{a} \in \mathcal{A}^n$ is the original joint action, and $\bm{\tilde{a}}_c$ is the most recent joint decision made by all agents $i_c$. Subsequently, they move to the next state $\widehat{s}'$ according to $\widehat{\mathcal{P}} (\widehat{s}' | \widehat{s}, \widehat{\bm{a}}) = \mathcal{O_P} (\bm{\tilde{o}}'_c | s') \mathcal{P} (s' | s, \bm{a})$ and earn a joint reward $\widehat{r} (\widehat{s}, \widehat{\bm{a}}) = r (s, \bm{a})$\footnote{This setup allows us to investigate the effect of asynchronous actions $\bm{\tilde{a}}_c$ on the reward $r(s, \bm{a})$.}.
\end{definition}

To ensure consistent dimensions of the extended state $\widehat{s}$ and action $\widehat{\bm{a}}$, virtual proxies are introduced for decision-making agents $i_d \in \mathcal{N}_d$ and action-executing agents $i_c \in \mathcal{N}_c$  at each time step, with proxies for agents $i_d$ masked out. 

Our VSP improves the training efficiency without increasing the complexity. Since a virtual proxy shares policy with the corresponding real agent, the introduction of a proxy does not expand the policy space. Furthermore, during the execution of action $a^1_3$ in Figure \ref{fig:adex}, virtual proxy $\#3'$ is repeatedly introduced, enabling multiple updates to the shared policy associated with action $a^1_3$. This significantly accelerates convergence, as demonstrated by the ablation studies in Section \ref{main:as}. Theoretically, we prove that VSP preserves the inherent characteristics of the task and the solution process.

\begin{theorem}
	\label{th:theorem1}
	Given an asynchronous decision-making task, define $\bm{\pi}_{Dec}^*$ and $\bm{\pi}_{VSP}^*$ as the Markov Perfect Equilibrium (MPE) respectively for modeling as a Dec-POMDP and a Dec-POMDP with VSP. $\mathcal{T}_{Dec}^*$ and $\mathcal{T}_{VSP}^*$ as the VD operator for the same. Assuming that $\mathcal{T}_{Dec}^*$ converges to the MPE of this task, i.e., $\bm{\pi}_{Dec}^*$, then: \\
	(1) $\bm{\pi}_{VSP}^* = \bm{\pi}_{Dec}^*$\footnote{Since VSP does not affect the policy space, the joint policy can still be denoted by $\bm{\pi}$ from Dec-POMDP with $n$ agents.}; \\
	(2) Given the same initial joint policy $\bm{\pi}_0$, $\mathcal{T}_{Dec}^*$ and $\mathcal{T}_{VSP}^*$ converge to the same MPE.
\end{theorem}

\begin{proof}
	Please see Appendix B.
\end{proof}

	\section{MVD}

\subsection{Multiplicative Interaction Form}

Based on VSP, asynchronous credit assignment can be addressed in a synchronized manner within a single time step. According to the unified framework of general VD algorithms, the global Q-value $Q_{tot}$ with VSP mechanism is formulated in terms of individual utilities $Q_i$ as follows:
\begin{equation}
	\label{eq:avd}
	Q_{tot} = Q_{tot}(s, Q_{1_d}, \cdots, Q_{n_d}, Q_{1_c}, \cdots, Q_{n_c}, Q_{1'_c}, \cdots, Q_{n'_c}).
\end{equation}

In asynchronous tasks, the agent who executes first must predict how later choices of other agents would affect its execution. Meanwhile, the agent who executes later needs to consider the impact of the current actions of other agents on its decision. This \textbf{conditional dependency} is represented in Eq. (\ref{eq:avd}) as the \textbf{interaction} between $Q_{i_d}$ and $Q_{i'_c}$. Considering the benefits of multiplicative interactions for fusing information streams and enabling conditional computation \cite{miold,mi}, we enrich Eq. (\ref{eq:ai}) by incorporating multiplicative interactions to capture these dependencies. We propose the general Multiplicative Value Decomposition (MVD) formula:
\begin{equation}
	\label{eq:mvd}
	Q_{tot} = k_0 + \sum_i^{n+n'_c} k_i Q_i + \sum_{i_d, i'_c} k_{i_di'_c} Q_{i_d} Q_{i'_c}.
\end{equation}

The detailed derivations are provided in Appendix C.1. We derive Eq. (\ref{eq:mvd}) by performing a Taylor expansion of $Q_i$ near an optimal joint action, providing support for the multiplication operation present in the value decomposition process.

Furthermore, compared to the traditional additive interaction VD, the multiplicative interaction between $Q_{i_d}$ and $Q_{i'_c}$ in MVD significantly boosts representational capability in learning interactions among agents. For Eq. (\ref{eq:ai}), the gradient for updating $Q_i$ is $\frac{\partial Q_{tot}}{\partial Q_i} = k_i$. In contrast, the gradients from Eq. (\ref{eq:mvd}) are $\frac{\partial Q_{tot}}{\partial Q_{i_d}} = k_{i_d} + \sum_{i'_c} k_{i_di'_c} Q_{i'_c}$, $\frac{\partial Q_{tot}}{\partial Q_{i'_c}} = k_{i'_c} + \sum_{i_d} k_{i_di'_c} Q_{i_d}$, and $\frac{\partial Q_{tot}}{\partial Q_{i_c}} = k_{i_c}$. Therefore, MVD integrates the nonlinear interactions as contextual information, allowing $Q_{i_d}$ and $Q_{i'_c}$ to refine their policies based on their mutual influence. We prove that MVD bears advantages in handling asynchronous tasks over traditional VD.

\begin{theorem}
	\label{th:theorem2}
	Given an asynchronous decision-making task, define $\mathcal{Q}_{Add}$ as the function class for the additive VD operator $\mathcal{T}_{Add}^*$, and $\mathcal{Q}_{Mul}$ as the function class for the multiplicative interaction VD operator $\mathcal{T}_{Mul}^*$, then:
	\begin{equation}
		\mathcal{Q}_{Add} \subsetneq \mathcal{Q}_{Mul}  \nonumber.
	\end{equation}
\end{theorem}

\begin{proof}
	Please see Appendix B.
\end{proof}

\subsection{High-Order Interaction Form}

Eq. (\ref{eq:mvd}) primarily considers the interaction of $Q_{i_d}$ and $Q_{i'_c}$. However, as illustrated in Figure \ref{fig:adex_1}, the actions $a_1^0$ and $a_3^1$ re-executed by virtual proxies $\#1'$ and $\#3'$ at $t_2$ actually originate from different time steps. This implies interactions between $Q_{1'}$ and $Q_{3'}$, thereby indicating high-order interactions among $Q_{1'}$, $Q_{3'}$, and $Q_2$. Therefore, we propose a $K$-th order (where $1 \le K \le n$) interactive VD formula as follows:
\begin{equation}
	\label{eq:high}
	\begin{aligned}
		&Q_{tot} = k_0 + \sum_i^{n+n'_c} k_i Q_i + \sum_{i_d, i'_c} k_{i_di'_c} Q_{i_d} Q_{i'_c} \\
		&+ \cdots + \sum_{i_d, i'_{c,1}, \cdots, i'_{c,K-1}} k_{i_di'_{c,1} \cdots i'_{c,K-1}} Q_{i_d} Q_{i'_{c,1}} \cdots Q_{i'_{c,K-1}}.
	\end{aligned}
\end{equation}

The detailed derivations are provided in Appendix C.2. Based on the derivation of Eq. (\ref{eq:high}), as the order increases, the error introduced by Taylor expansion decreases and agents are able to obtain deeper interactive information. Nevertheless, higher-order interaction complicates the model and does not necessarily lead to better final performance \cite{pr2,na2q}. Our ablation studies in Section \ref{main:as} further confirm this conclusion. Therefore, we primarily focus on multiplicative interactions between $Q_{i_d}$ and $Q_{i'_c}$.

\begin{figure*}[!t]
	\centering
	\includegraphics[width=0.86\textwidth]{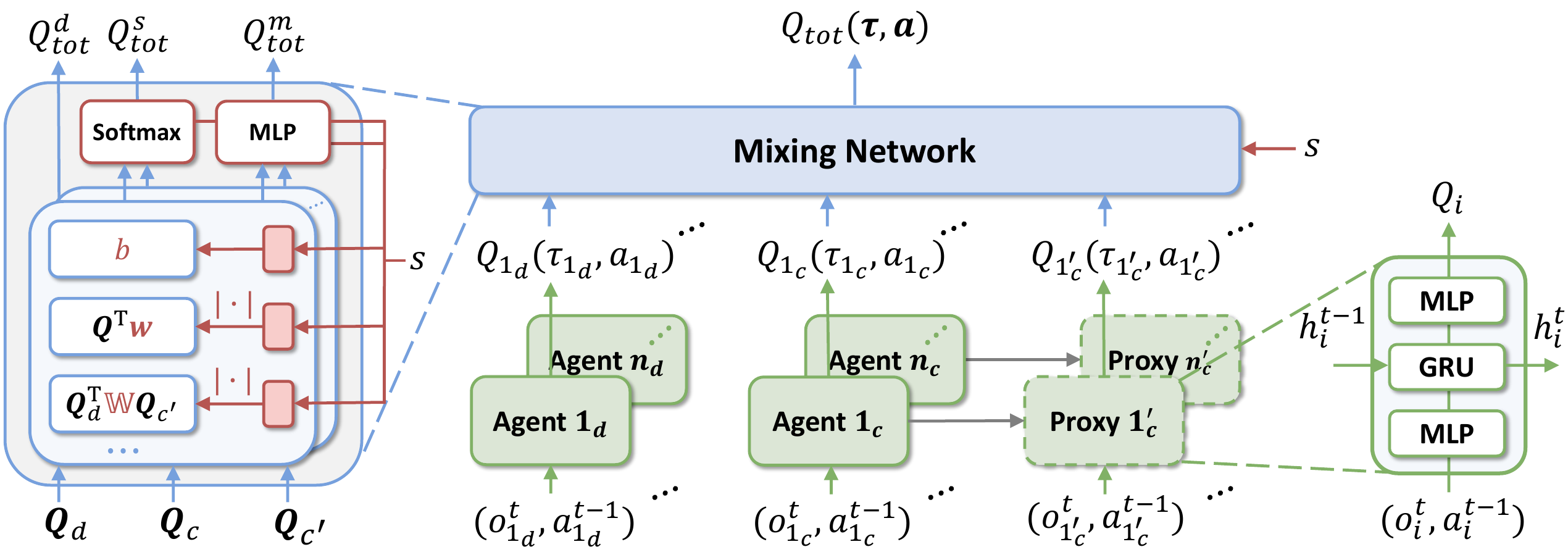}
	\caption{The overall framework of MVD. \textbf{Left:} Mixing network structure. In red are the hypernetworks that generate the weights and biases for mixing network. \textbf{Middle:} The overall MVD architecture. \textbf{Right:} Agent network structure.}
	\label{fig:mvd}
\end{figure*}

\subsection{Implementation}

Finally, we discuss the issue of IGM consistency in the practical implementation of MVD. In Dec-POMDP with VSP, the agent $i_c$ currently executing an action does not need to choose a new one, and virtual proxy $i'_c$ can only execute the asynchronous actions of $i_c$. This implies that agent $i_c$ and proxy $i'_c$ do not need to satisfy the IGM condition. Consequently, we obtain the MVD-based IGM as follows:
\begin{equation}
	\begin{aligned}
		\label{eq:mvdigm}
		\arg\max_{\bm{a}}Q_{tot}(s,\bm{a}) =
		(&\arg\max_{a_{1_d}} Q_{1_d}(\tau_{1_d}, a_{1_d}), \\
		&\cdots, \arg\max_{a_{n_d}} Q_{n_d}(\tau_{n_d}, a_{n_d}), \\
		&a_{1_c}, \cdots, a_{n_c}, a_{1'_c}, \cdots, a_{n'_c}).
	\end{aligned}
\end{equation}

To maintain the monotonicity between $Q_{tot}$ and $Q_{i_d}$, i.e., $\frac{\partial Q_{tot}}{\partial Q_{i_d}} \ge 0$, we derive the following form that satisfies MVD-based IGM and employ hypernetworks \cite{hyper} $f_i(s)$ to learn the corresponding weights. Since $Q_{i'_c}$ may be less than 0, we track the minimum $Q_{i'_c}$ during training and denote it as $Q^{min}_c$, ensuring $Q_{i'_c} + Q^{min}_c \ge 0$. The detailed derivations are provided in Appendix C.3.
\begin{equation}
	\label{eq:mvd2}
	Q_{tot} \approx f_0 + \sum_i^{n+n'_c} |f_i| Q_i + \sum_{i_d, i'_c} |f_{i_di'_c}| \frac{Q_{i'_c} + Q^{min}_c}{2} Q_{i_d}.
\end{equation}

The overall framework of MVD is illustrated in Figure \ref{fig:mvd}. We propose three distinct practical implementations of the mixing network. The first approach directly calculates the final global Q value based on Eq. (\ref{eq:mvd2}), denoted as $Q_{tot}^d$. The second approach employs a multi-head structure that allows the mixing network to focus on asynchronous interaction information from different representation sub-spaces, thereby enhancing the representational capability and stability. Each head calculates a global Q-value based on Eq. (\ref{eq:mvd2}), and inputs it into a Softmax function to obtain the final global Q-value, denoted as $Q_{tot}^s$. The third approach also employs a multi-head mechanism. To simplify the model, we use a ReLU activation function and an MLP to replace the Softmax function in the second implementation, denoted as $Q_{tot}^m$. In this paper, we primarily focus on the third implementation, and comparisons with the others are discussed in Section \ref{main:as}. The pseudo-code for MVD is in Appendix D.

	\section{Experiments}

In this section, we evaluate our MVD on a modified asynchronous variant of SMAC and two existing asynchronous benchmarks: Overcooked and POAC. SMAC is a widely used testbed for MARL algorithms. Our asynchronous variant introduces asynchronous actions, where allied agents require different time steps to complete their movements. Overcooked requires agents to prepare ingredients in sequence and cooperate to make salads. Different actions, such as chopping, moving to ingredients, and delivering, span varying time steps. POAC is a confrontation wargame between two armies with three unit types, each having distinct attributes and action execution times. The goal is to learn asynchronous strategies to defeat the rule-based bots.

The baselines fall into three categories: (1) The decentralized training and decentralized execution (DTDE) method, IPPO \cite{ippo}, treats other agents as part of the environment and is applicable to asynchronous tasks. (2) The discarding type method, MAC IAICC, which is the most advanced algorithm in \cite{maciaicc}. (3) Credit assignment algorithms based on padding type method, including QMIX, Qatten, SHAQ \cite{shaq}, ICES \cite{ices}, and NA$^2$Q that considers 2nd-order interactions\footnote{We excluded other asynchronous MARL algorithms in the section of Related Works due to their specificity to tasks with individual agent rewards or lack of open-source code.}.

Details of benchmarks, baselines, and our MVD are provided in Appendix E. The graphs illustrate the performance of all compared algorithms by plotting the mean and standard deviation of results obtained across five random seeds.

\begin{figure}[t]
	\centering
	\includegraphics[width=0.48\textwidth]{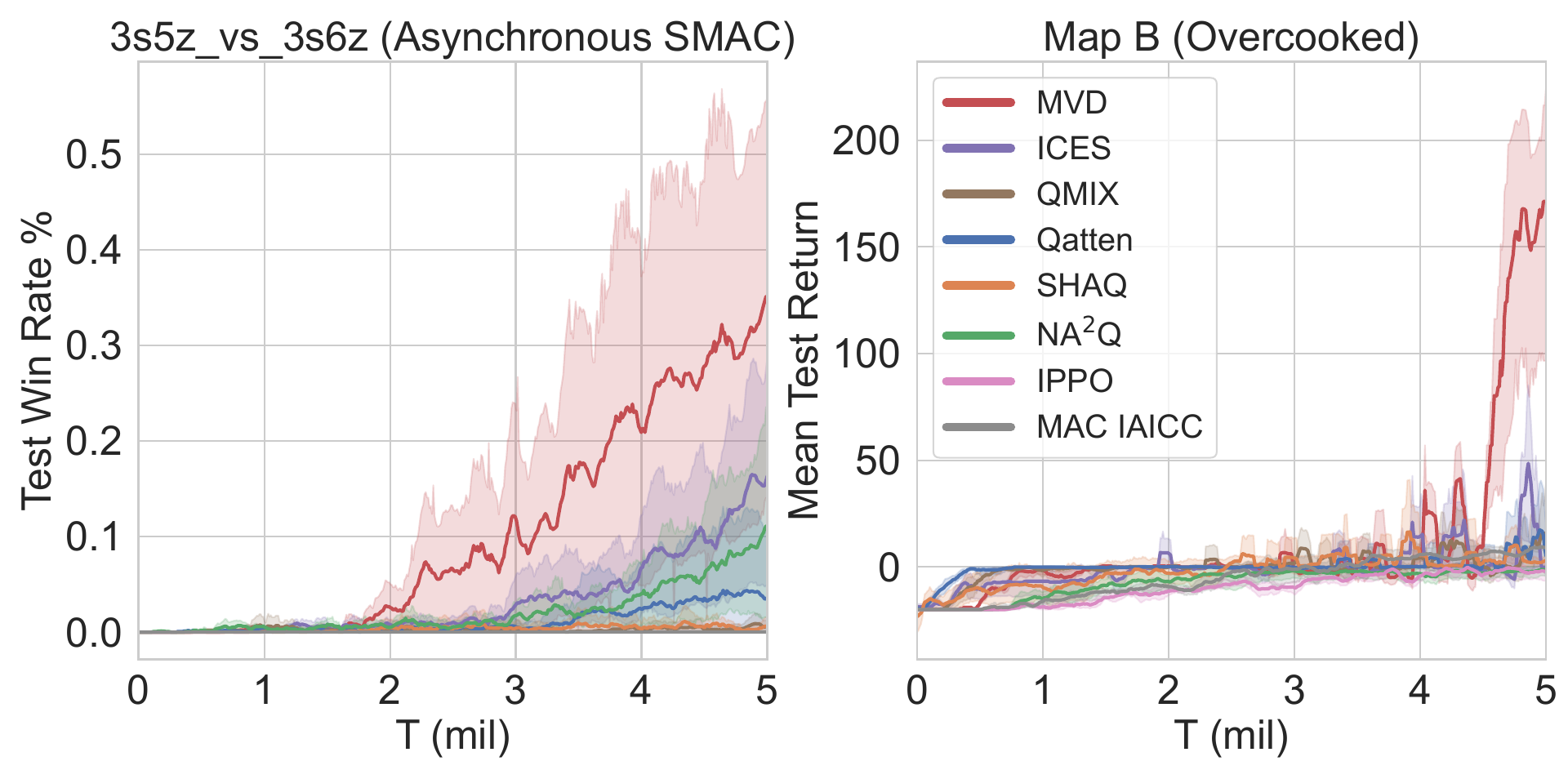}
	\caption{Performance on two challenging asynchronous scenarios}
	\label{fig:hard}
\end{figure}

\begin{figure*}[t]
\centering
\includegraphics[width=0.96\textwidth]{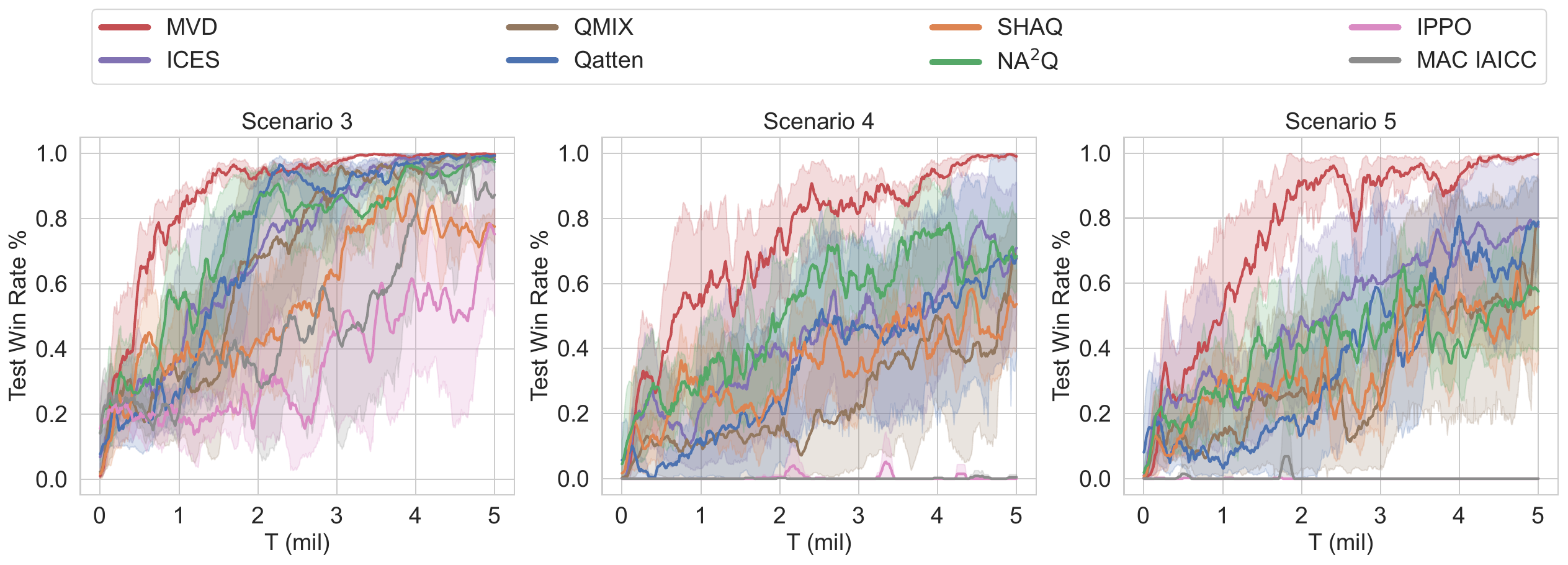}
\caption{Test win rate \% on three scenarios of POAC benchmark.}
\label{fig:bq}
\end{figure*}

\begin{figure}[t]
	\centering
	\includegraphics[width=0.45\textwidth]{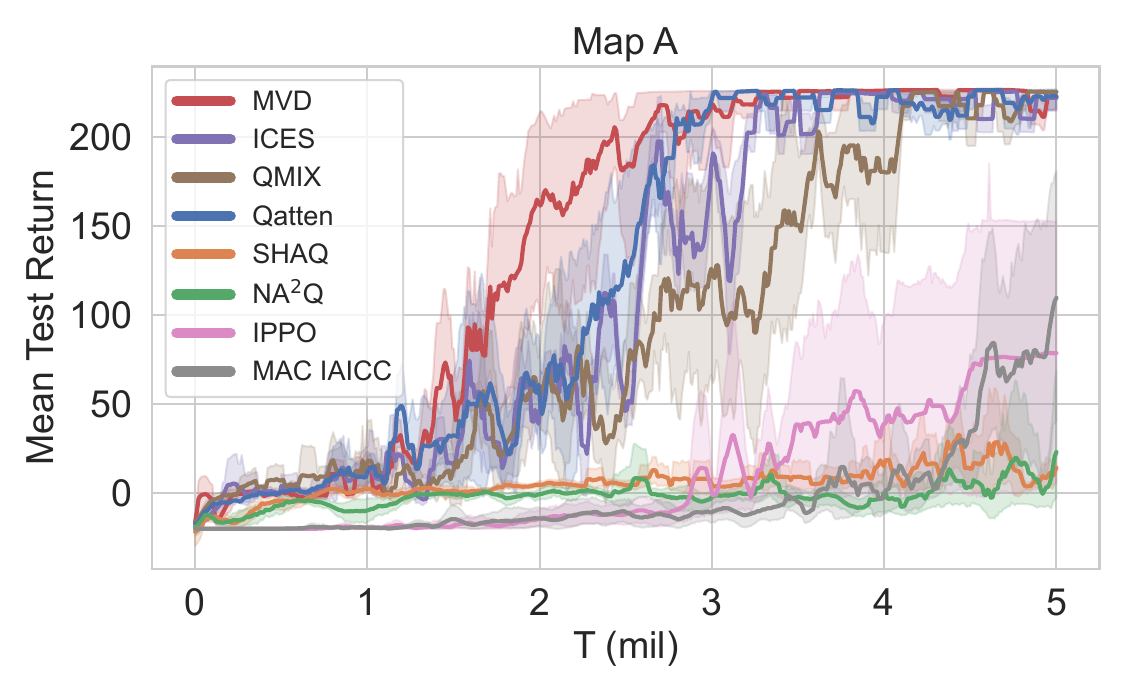}
	\caption{Mean test return on Overcooked benchmark.}
	\label{fig:oc_a}
\end{figure}

\subsection{Benchmark Results}

We run experiments across multiple benchmarks, focusing on three key aspects of our framework in asynchronous cooperation: the necessity of additional computation, effectiveness against baselines, and generalization in complex tasks.

We first investigate why the additional computations introduced by VSP and MVD are necessary for asynchronous cooperation. As shown in Figure \ref{fig:hard}, in the super hard scenario of asynchronous SMAC and map B of Overcooked, both discarding and padding methods fail to handle asynchronous cooperation effectively. In contrast, MVD employs virtual proxies and multiplicative interactions to better capture the marginal contributions of asynchronous actions, significantly accelerating convergence. The complete experimental results and analysis are provided in Appendix F. These results show that while most baselines perform well in simple asynchronous scenarios, they struggle in complex tasks.

We then analyze the specific performance comparison against baselines on map A of Overcooked, as shown in Figure \ref{fig:oc_a}. The results indicate that MVD surpasses all baselines. Both IPPO and MAC IAICC exhibit slower training speeds. This suggests the discarding type methods suffer from low efficiency. NA$^2$Q and SHAQ mistakenly consider the influence among padding actions, resulting in non-convergence. This implies that Dec-POMDP with padding action is also unsuitable for asynchronous settings. Although ICES enhances exploration, it is less effective than MVD as it neglects the interplay between asynchronous actions. QMIX and Qatten perform better than NA$^2$Q because they use simpler models to handle credit assignment, leading to stronger robustness to padding action. Furthermore, MVD outperforms baselines on other maps of Overcooked. The complete experimental results and analysis are in Appendix F.2.

We further validate the generalization of MVD on the challenging POAC. Figure \ref{fig:bq} shows the win rate across scenarios with increasing difficulty levels. We observe that MVD demonstrates increasingly better performance. IPPO and MAC IAICC train slowly in simple scenarios and fail in complex tasks. Compared to Overcooked, POAC contains fewer asynchronous actions, resulting in less padding when using Dec-POMDP. Therefore, NA$^2$Q performs relatively better among the baselines. However, due to the interference from padding action and the complexity of the adopted model, the training efficiency of NA$^2$Q is consistently inferior to that of MVD, a result also observed in SHAQ and ICES. The additive interaction VD algorithms, QMIX and Qatten, do not yield satisfactory performance, since they cannot handle the mutual influence among agents in complex asynchronous tasks. Furthermore, MVD demonstrates highly competitive performance in other scenarios of POAC. The complete experimental results and analysis are in Appendix F.3.

\subsection{Ablation Studies}
\label{main:as}

To obtain a deeper insight into our proposed VSP and MVD, we perform ablation studies to illustrate the impact of the following factors on the performance: (1) The introduction of virtual proxies. (2) The interactions of different orders among agents. (3) The distinct practical implementations of MVD.

\begin{figure*}[t]
	\centering
	\subfloat[Virtual proxies]{
		\includegraphics[width=0.32\textwidth]{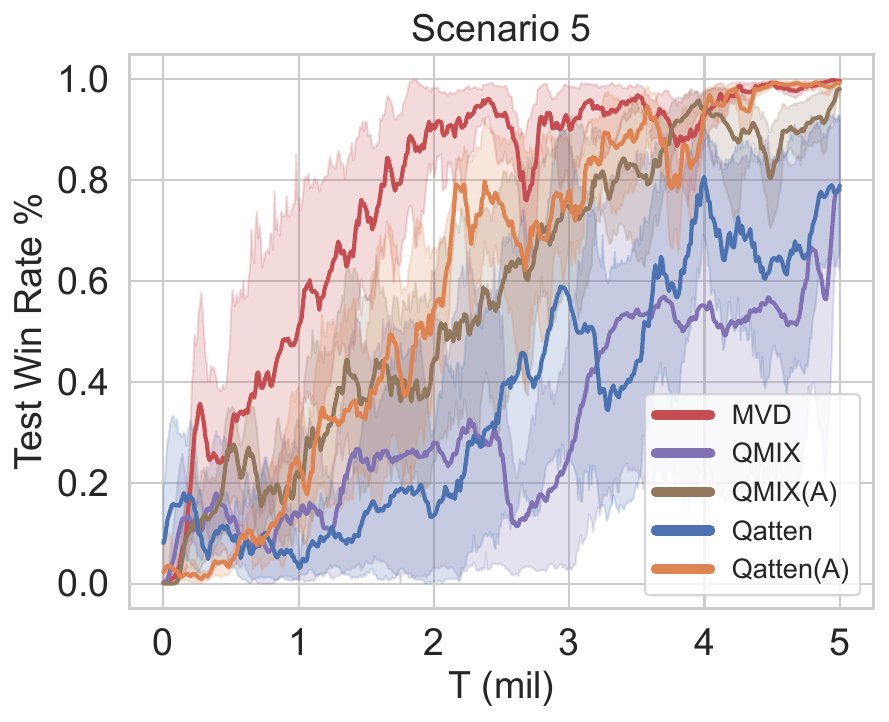}
		\label{fig:bq_adex}
	}
	\hfil
	\subfloat[Different orders interactions]{
		\includegraphics[width=0.32\textwidth]{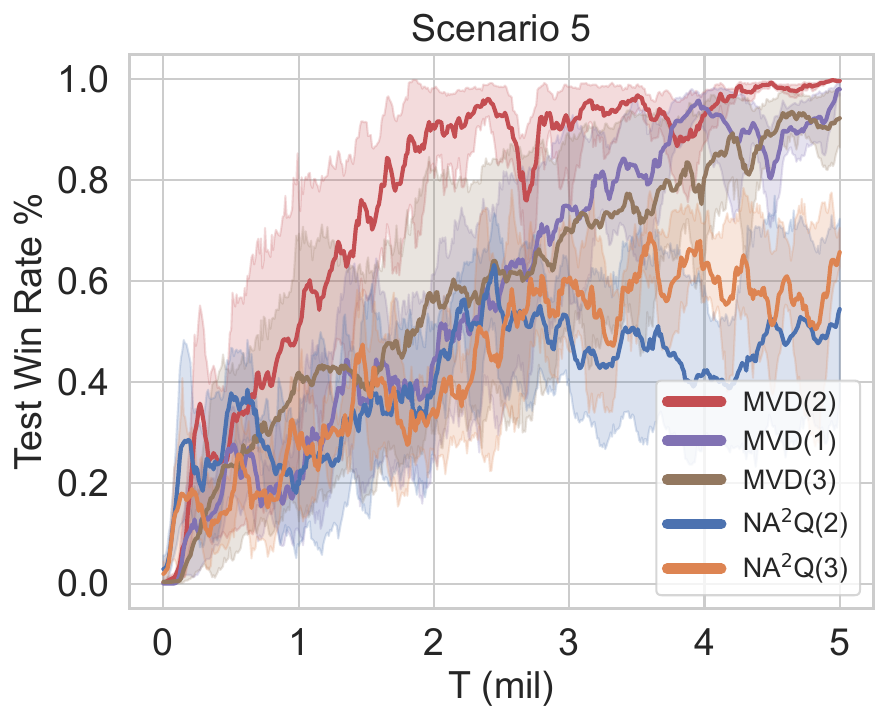}
		\label{fig:bq_high}
	}
	\hfil
	\subfloat[Different implementations]{
		\includegraphics[width=0.32\textwidth]{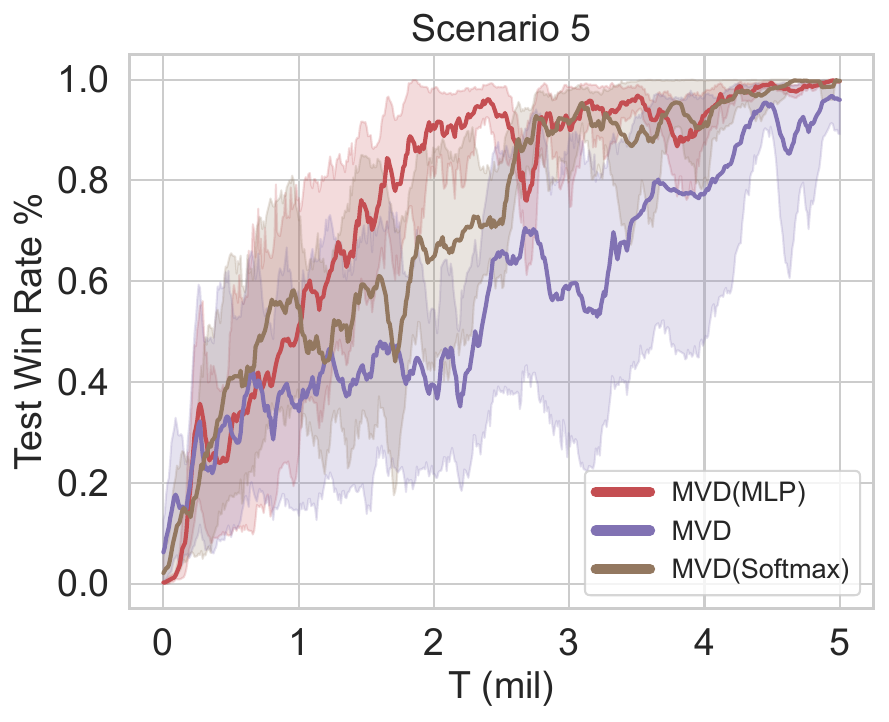}
		\label{fig:bq_f}
	}
	\caption{Ablation Studies of MVD on scenario 5 of POAC benchmark.}
	\label{fig:ablation_study}
\end{figure*}

For evaluating the impact of factor (1), we extend QMIX and Qatten with VSP, denoting them as QMIX(A) and Qatten(A). As shown in Figure \ref{fig:bq_adex}, the introduction of virtual proxies does not complicate the problem but instead notably improves the performance of QMIX and Qatten, highlighting the powerful advantages of VSP in an asynchronous setting. However, they fail to capture the mutual influence among agents, leading to poorer performance than MVD. We also extend other VD algorithms with VSP. The complete ablation experiments and analysis are in Appendix G.1.

For evaluating the impact of factor (2), we consider MVD and NA$^2$Q with $l$-th order interactions under VSP, denoting them as MVD($l$) and NA$^2$Q($l$). As shown in Figure \ref{fig:bq_high}, the performance of MVD(2) incorporating multiplicative interactions consistently outperforms MVD(1) which does not consider interactions between agents. Further, both MVD(3) and NA$^2$Q(3) suffer from performance degradation due to increased model complexity. Therefore, additive interactions, which ignore interactions, are insufficient for solving the asynchronous credit assignment problem, while higher-order interactions increase complexity and degrade performance. In contrast, the multiplicative interaction between $Q_{i_d}$ and $Q_{i'_c}$ efficiently addresses these issues. The complete ablation experiments and analysis are in Appendix G.2.

For evaluating the impact of factor (3), we compared three different practical implementations of MVD. As shown in Figure \ref{fig:bq_f}, directly applying Eq. (\ref{eq:mvd2}) to obtain $Q_{tot}^d$ converges to the optimal joint policy, yet it suffers from slow training speed and instability. Employing a multi-head structure can effectively address these issues. However, using Softmax to obtain $Q_{tot}^s$ excessively complicates the entire mixing network. Therefore, MVD derives the greatest benefit from MLP with the ReLU activation function. The complete ablation experiments and analysis are in Appendix G.3.

\subsection{Interpretability}

\begin{figure}[t]
	\centering
	\subfloat[]{
		\includegraphics[width=0.24\textwidth]{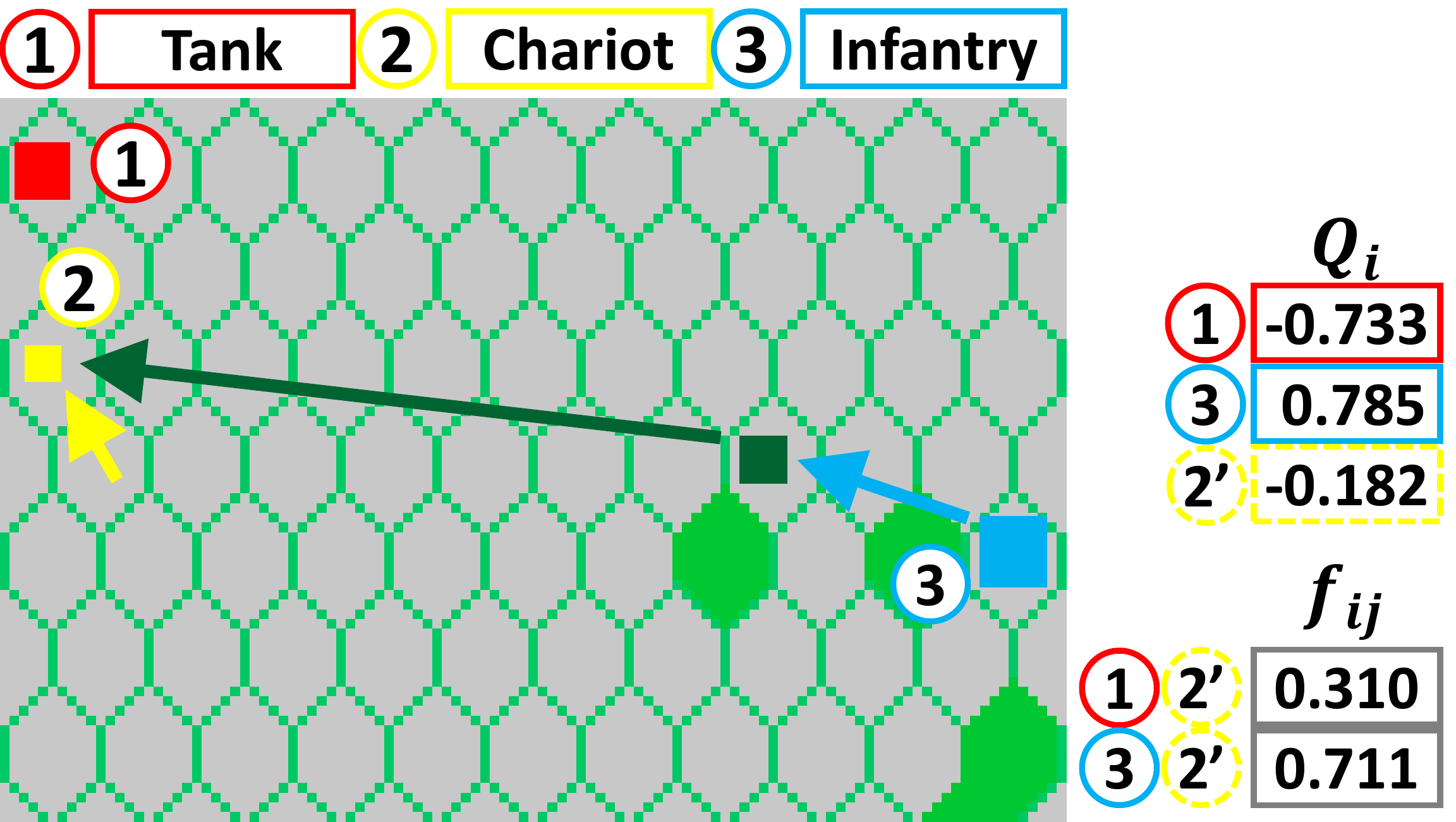}
		\label{fig:visual_1}
	}
	\subfloat[]{
		\includegraphics[width=0.24\textwidth]{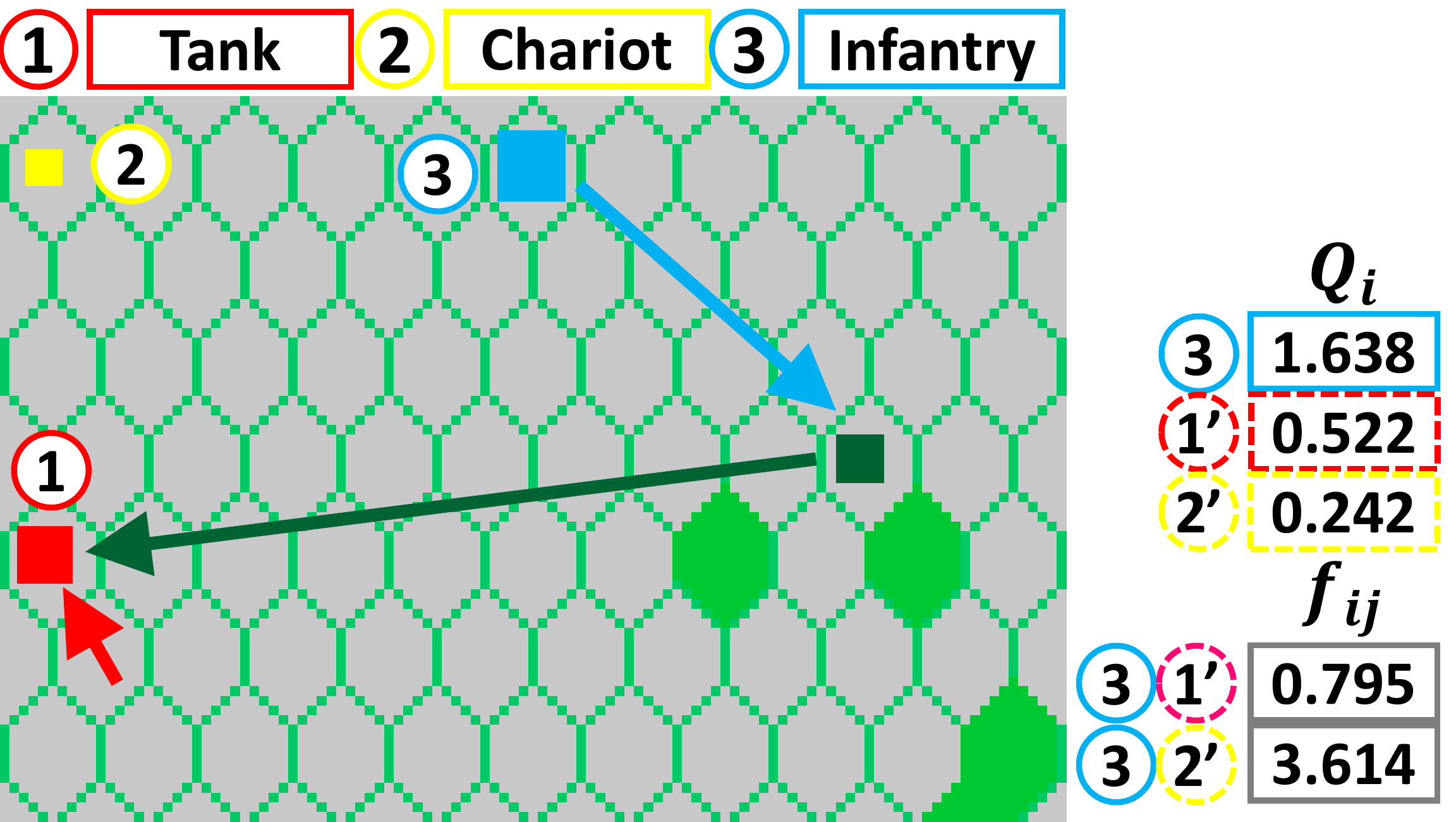}
		\label{fig:visual_2}
	}
	\caption{Visualization of evaluation for (a) MVD and (b) NA$^2$Q.}
\end{figure}

To visually illustrate the asynchronous credit assignment process, we exhibit key frames from scenario 5 of POAC and compare the converged MVD and NA$^2$Q with VSP. The arrows in the figures represent the movements or attacks of the agents. We highlight the individual $Q_i$ and crucial weights within the mixing network to demonstrate their alignment with agents' asynchronous behaviors. $f_{ij}$ represents the collaborative contribution of two agents to the global outcome.

Figures \ref{fig:visual_1} and \ref{fig:visual_2} depict similar scenarios: one of our units lures the enemy deeper into the field, while another attack. Since the former is executing a movement action and the latter is making an attack decision, their decision-making is asynchronous. As shown in Figure \ref{fig:visual_1}, the infantry successfully attacks an enemy, leading to a higher $Q_3$, while the chariot is under attack, resulting in a negative $Q_{2'}$. MVD accurately attributes the importance of their asynchronous cooperation to the entire team using multiplicative interaction and assigns a higher credit $f_{32'}$. As shown in Figure \ref{fig:visual_2}, similarly, the infantry that successfully attacks enemies has a higher $Q_3$, whereas the tank used to kite enemies has a lower $Q_{1'}$. Both the tank and the chariot are executing actions, yet NA$^2$Q mistakenly regards the asynchronous infantry-chariot cooperation as more important than infantry-tank cooperation. Therefore, even though both strategies ultimately achieve victory, MVD offers a superior ability to capture the interplay between agents' asynchronous decision-making, providing higher interpretability.

	\section{Conclusion and Future Work}

In this paper, we propose an asynchronous credit assignment framework incorporating the VSP mechanism and the MVD algorithm. Our framework fully captures the dependencies between asynchronous decisions and provides a solid basis for further exploration of asynchronous MARL. VSP synchronizes asynchronous actions without disrupting task equilibrium or VD convergence. MVD introduces multiplicative interactions, strictly extending the function class and effectively capturing the interplay between asynchronous actions. Extensive experiments demonstrate that MVD outperforms baselines, particularly in complex asynchronous tasks, and provides interpretability for asynchronous cooperation. One direction for future work involves exploring effective representations of higher-order asynchronous interactions and addressing asynchronous cooperation in large-scale systems.


\section*{Acknowledgments}
This paper was mainly supported by the National Natural Science Foundation of China (NSFC) under Grant 62272497 to H. Wu.

	\bibliographystyle{named}
	\bibliography{10_reference}
	
	\newpage
	\clearpage

	\newpage
	\appendix
	\onecolumn
	
	\section{Credit Assignment Methods}
\label{app:ca}

The centralized training and decentralized execution (CTDE) \cite{ctde} framework is widely adopted by modern MARL methods. It leverages global information for centralized training, enabling agents to make decentralized decisions based on their local information. A key challenge in CTDE is the credit assignment problem, which investigates agents' marginal contributions to the overall success, facilitating better learning of agent policies and enhancing cooperation among agents \cite{review}.

Value decomposition (VD) stands as one of the most prevalent implicit credit assignment methods \cite{lica}, which learns the marginal contribution of each agent and decomposes the global Q-value function $Q_{tot}$ into individual agent utility functions $Q_i$. VDN \cite{vdn} directly assumes equal contributions from all agents and decomposes $Q_{tot}$ into the sum of $Q_i$s. QMIX utilizes a mixing network to learn the nonlinear relationship between $Q_{tot}$ and $Q_i$. Since QMIX constructs the mixing network using MLP layers, we still consider QMIX as the additive interaction form \cite{mi}. QPLEX \cite{qplex} introduces the dueling structure $Q = V + A$ \cite{dueling} and performs VD on the global state value function $V_{tot}$ as well as the global advantage value function $A_{tot}$, respectively. Qatten theoretically derives a general additive interaction formula of $Q_{tot}$ in terms of $Q_i$ and proposes practical multi-head attention to approximate the weight of $Q_i$. DVD \cite{dvd} employs backdoor adjustment to remove the confounding bias in VD, aiming to obtain better weights of $Q_i$. In summary, the VD algorithms mentioned above can be formulated as the additive interaction form as Eq. (\ref{eq:ai}), which leads to limitations in addressing the underlying dependencies among agents \cite{qmix,na2q}. Therefore, NA$^2$Q \cite{na2q} introduces a generalized additive model (GAM) \cite{gam} to model all possible higher-order interactions among agents and to interpret their collaboration behavior.

On the other hand, explicit credit assignment methods \cite{lica} introduce different concepts to explicitly calculate the marginal contribution of each agent. Inspired by difference rewards \cite{difference}, COMA \cite{coma} proposes a counterfactual baseline to calculate the specific critic network for each agent. QPD \cite{qpd} employs integrated gradients \cite{integrated} to attribute the $Q_{tot}$ to each agent’s feature changes. SQPPDG \cite{sqddpg} and SHAQ \cite{shaq} extend shapley value \cite{shapley} for the credit assignment in an extended convex game with a grand coalition. ICES \cite{ices} employs the Bayesian surprise to characterize individual contributions, guiding agents in their exploration. However, compared with explicit credit assignment, implicit methods offer superior generalization capabilities and can autonomously learn the marginal contributions of agents \cite{lica}. Therefore, in this paper, we choose to extend the implicit credit assignment method, VD, to the asynchronous domain.

	\section{Proofs}  
\label{app:proofs}

\begin{definition}
	\label{def:mpe}
	A joint policy $\bm{\pi}^* = \{ \pi_i, \pi_{-i} \}$ is called a Markov Perfect Equilibrium (MPE) if:
	\begin{equation*}
		V^{\pi_i, \pi_{-i}}(s) \ge V^{\tilde{\pi}_i, \pi_{-i}}(s), \ \forall s \in \mathcal{S},\ i \in \mathcal{N},\ \forall \tilde{\pi}_i.
	\end{equation*}
\end{definition}

\begin{definition}
	\label{def:boo}
	The Bellman optimality operator $\mathcal{T}^*$ for Q-learning variants in MARL is:
	\begin{equation*}
		\mathcal{T}^*Q(s,\bm{a}) = \mathbb{E}_{s' \sim \mathcal{P}}[r + \gamma \max_{\bm{a}'} Q(s', \bm{a}')].
	\end{equation*}
\end{definition}

\begin{definition}
	\label{def:vdc}
	The function class $\mathcal{Q}_{tot}$ of the approximate global Q-value function $Q_{tot}$ in VD is:
	\begin{equation*}
		\mathcal{Q}_{tot} = \{ Q_{tot} | Q_{tot}(s,\bm{a}) = f_s(Q_1(\tau_1, a_1), \cdots, Q_n(\tau_n, a_n)),\text{\ where\ } Q_{tot} \text{\ and\ } Q_i \text{\ satisfy IGM} \}.
	\end{equation*}
\end{definition}

\begin{definition}
	\label{def:vdo}
	The VD operator is $\mathcal{T}^*_{VD} := \prod_{VD} \cdot \mathcal{T}^*$. $\mathcal{T}^*$ is the Bellman optimality operator to learn the ground truth global Q-value function $Q^*$. $\prod\nolimits_{VD}$ is the operator that finds the global Q-value function $Q_{tot}$ in the function class $\mathcal{Q}_{tot}$ to approximate $Q^*$, which is defined as:
	\begin{equation*}
		\prod\nolimits_{VD} Q(s,\bm{a}) = \mathop{\arg\min}_{q \in \mathcal{Q}_{tot}} (Q(s,\bm{a}) - q(s,\bm{a}))^2.
	\end{equation*}
\end{definition}

\begin{appendix_lemma}
	\label{lemma1}
	Given an asynchronous decision-making task and a joint policy $\bm{\pi}$, let $Q^{\bm{\pi}}_{Dec}$ and $Q^{\bm{\pi}}_{VSP}$ represent the global Q-value function converged by modeling the task with Dec-POMDP and Dec-POMDP with VSP, respectively. Then we have $Q^{\bm{\pi}}_{Dec} (s, \bm{a}) = Q^{\bm{\pi}}_{VSP} (\widehat{s}, \widehat{\bm{a}})$.
\end{appendix_lemma}

\begin{proof}
	We use the set $\mathcal{X}_{s}$ to represent different $\widehat{s}$ in Dec-POMDP with VSP that correspond to the same original $s$, and the set $\mathcal{Y}_{(s,\bm{a})}$ to represent different $(\widehat{s}, \widehat{\bm{a}})$ that correspond to the same original $(s,\bm{a})$. As specified in Definition \ref{def:adexpomdp}, $\widehat{s} = [s ; \bm{\tilde{o}}_c]$, and $\widehat{\bm{a}} = [\bm{a}; \bm{\tilde{a}}_c]$. For any $s$ and $s'$, we have $\mathcal{X}_{s} \cap \mathcal{X}_{s'} = \emptyset$. Similarly, we also have $\mathcal{Y}_{(s,\bm{a})} \cap \mathcal{Y}_{(s',\bm{a}')} = \emptyset$. Therefore, there is a one-to-one correspondence between $s$ and $\mathcal{X}_{s}$, as well as between $(s,\bm{a})$ and $\mathcal{Y}_{(s,\bm{a})}$, which implies that $\widehat{\mathcal{P}} (\mathcal{X}_{s'} | \mathcal{Y}_{(s,\bm{a})}) \equiv \mathcal{P} (s' | s, \bm{a})$ and $\widehat{r} (\mathcal{Y}_{(s,\bm{a})}) \equiv r (s, \bm{a})$.
	
	Given a joint policy $\bm{\pi}$, the global Q-value function converged in Dec-POMDP is defined as:
	\begin{equation*}
			Q^{\bm{\pi}}_{Dec} (s,\bm{a}) = \mathbb{E}_{(s,a) \sim (\mathcal{P}, \bm{\pi})} [r^1 + \gamma r^2 + \cdots + \gamma^{T-1} r^{T}].
	\end{equation*}

	And the global Q-value function converged in Dec-POMDP with VSP is defined as:
	\begin{align}
		Q^{\bm{\pi}}_{VSP} (\widehat{s}, \widehat{\bm{a}})
		&= \mathbb{E}_{(\widehat{s}, \widehat{\bm{a}}) \sim (\widehat{\mathcal{P}}, \bm{\pi})} [\widehat{r}^1 + \gamma \widehat{r}^2 + \cdots + \gamma^{T-1} \widehat{r}^{T}] \nonumber \\
		&= \mathbb{E}_{\mathcal{Y}_{(s,\bm{a})} \sim (\widehat{\mathcal{P}}, \bm{\pi})} \mathbb{E}_{(\bm{\tilde{o}}_c, \bm{\tilde{a}}_c) \sim (\mathcal{O_P}, \bm{\pi})} [\widehat{r}^1 + \gamma \widehat{r}^2 + \cdots + \gamma^{T-1} \widehat{r}^{T}] \nonumber \\
		&= \mathbb{E}_{\mathcal{Y}_{(s,\bm{a})} \sim (\widehat{\mathcal{P}}, \bm{\pi})} [ \mathbb{E}_{(\bm{\tilde{o}}_c, \bm{\tilde{a}}_c) \sim (\mathcal{O_P}, \bm{\pi})}[\widehat{r}^1] + \cdots + \gamma^{T-1} \mathbb{E}_{(\bm{\tilde{o}}_c, \bm{a}^{T_d}_{c}) \sim (\mathcal{O_P}, \bm{\pi})}[\widehat{r}^{T}]] \nonumber \\
		&= \mathbb{E}_{\mathcal{Y}_{(s,\bm{a})} \sim (\widehat{\mathcal{P}}, \bm{\pi})} [\widehat{r}^1 + \gamma \widehat{r}^2 + \cdots + \gamma^{T-1} \widehat{r}^{T}] \label{eq:explain_lemma1} \\
		&= \mathbb{E}_{(s,\bm{a}) \sim (\mathcal{P}, \bm{\pi})} [r^1 + \gamma r^2 + \cdots + \gamma^{T-1} r^{T}], \nonumber
	\end{align}
	where (\ref{eq:explain_lemma1}) represents the expectation over different cases in $\mathcal{Y}_{(s,\bm{a})}$. However, since $\widehat{r} (\mathcal{Y}_{(s,\bm{a})}) \equiv r (s, \bm{a})$, meaning that different cases in this set correspond to the same reward, the expectation operation can be removed. Finally, we prove that given the same $\bm{\pi}$, $Q^{\bm{\pi}}_{Dec} (s, \bm{a}) = Q^{\bm{\pi}}_{VSP} (\widehat{s}, \widehat{\bm{a}})$.
\end{proof}

\begin{appendix_lemma}
	\label{lemma2}
	Given an asynchronous decision-making task, we denote the Bellman optimality operator in Dec-POMDP as $\mathcal{T}^*_{Dec}$ and in Dec-POMDP with VSP as $\mathcal{T}^*_{VSP}$. Similarly, $Q^*_{Dec}$ and $Q^*_{VSP}$ represent the ground truth global Q-value functions when modeling the task using Dec-POMDP and Dec-POMDP with VSP, respectively. Then we have $Q^*_{Dec} (s, \bm{a}) = Q^*_{VSP} (\widehat{s}, \widehat{\bm{a}})$.
\end{appendix_lemma}

\begin{proof}
	Based on Lemma \ref{lemma1}, we have
	\begin{align}
		\mathcal{T}^*_{VSP} Q^{\bm{\pi}}_{VSP} (\widehat{s}, \widehat{\bm{a}}) 
		&= \mathbb{E}_{\widehat{s}' \sim \widehat{\mathcal{P}}}[\widehat{r} + \gamma \max_{\widehat{\bm{a}}'} Q^{\bm{\pi}}_{VSP} (\widehat{s}', \widehat{\bm{a}}')] \nonumber \\
		&= \mathbb{E}_{\mathcal{X}_{s'} \sim \widehat{\mathcal{P}}} \mathbb{E}_{\bm{\tilde{o}}_c \sim \mathcal{O_P}} [\widehat{r} + \gamma \max_{\widehat{\bm{a}}'} Q^{\bm{\pi}}_{VSP} (\widehat{s}', \widehat{\bm{a}}')] \nonumber \\
		&= \mathbb{E}_{\mathcal{X}_{s'} \sim \widehat{\mathcal{P}}} [\widehat{r} + \gamma \max_{\widehat{\bm{a}}'} Q^{\bm{\pi}}_{VSP} (\widehat{s}', \widehat{\bm{a}}')] \label{eq:explain_lemma2} \\
		&= \mathbb{E}_{s' \sim \mathcal{P}} [r + \gamma \max_{\bm{a}'} Q^{\bm{\pi}}_{Dec} (s', \bm{a}')] \nonumber \\
		&= \mathcal{T}^*_{Dec} Q^{\bm{\pi}}_{Dec} (s, \bm{a}), \nonumber
	\end{align}
	where (\ref{eq:explain_lemma2}) is derived based on the fact that for different $\widehat{s}_1$ and $\widehat{s}_2$ in $\mathcal{X}_{s}$, we have $Q^{\bm{\pi}}_{VSP}(\widehat{s}_1, \widehat{\bm{a}}) = Q^{\bm{\pi}}_{VSP}(\widehat{s}_2, \widehat{\bm{a}}) = Q^{\bm{\pi}}_{Dec}(s, \bm{a})$, thus allowing us to eliminate the expectation operation. Hence, assuming the same initial joint policy $\bm{\pi}_0$, both $\mathcal{T}^*_{Dec}$ and $\mathcal{T}^*_{VSP}$ can converge to the same optimal global Q-value function, implying $Q^*_{Dec} (s, \bm{a}) = Q^*_{VSP} (\widehat{s}, \widehat{\bm{a}})$.
\end{proof}

\begin{appendix_theorem}
	Given an asynchronous decision-making task, define $\bm{\pi}_{Dec}^*$ and $\bm{\pi}_{VSP}^*$ as the Markov Perfect Equilibrium (MPE) respectively for modeling as a Dec-POMDP and a Dec-POMDP with VSP. $\mathcal{T}_{Dec}^*$ and $\mathcal{T}_{VSP}^*$ as the VD operator for the same. Assuming that $\mathcal{T}_{Dec}^*$ converges to the MPE of this task, i.e., $\bm{\pi}_{Dec}^*$, then: \\
	(1) $\bm{\pi}_{VSP}^* = \bm{\pi}_{Dec}^*$; \\
	(2) Given the same initial joint policy $\bm{\pi}_0$, $\mathcal{T}_{Dec}^*$ and $\mathcal{T}_{VSP}^*$ converge to the same MPE.
\end{appendix_theorem}

\begin{proof}
	Firstly, We prove that Dec-POMDP with VSP preserves the original MPE mentioned in Definition \ref{def:mpe}. According to Lemma \ref{lemma1}, given any joint policy $\bm{\pi}$, the converged global Q-value function $Q^{\bm{\pi}}_{Dec}$ under Dec-POMDP is equal to the converged global Q-value function $Q^{\bm{\pi}}_{VSP}$ under Dec-POMDP with VSP. Thus, we have $Q^{\bm{\pi}}_{Dec} (s, \bm{a}) = Q^{\bm{\pi}}_{VSP} (\widehat{s}, \widehat{\bm{a}})$, which implies $V^{\bm{\pi}}_{Dec} (s) = V^{\bm{\pi}}_{VSP} (\widehat{s})$. Therefore, the original MPE $\bm{\pi}^*$ in Dec-POMDP is also an MPE in Dec-POMDP with VSP.
	
	Secondly, regarding the VD operator $\mathcal{T}^*_{VD} = \prod\nolimits_{VD} \cdot \mathcal{T}^*$ mentioned in Definition \ref{def:vdo}. On the one hand, according to Lemma \ref{lemma2}, the introduction of virtual proxies does not affect $\mathcal{T}^*$, which enables learning of the same $Q^*$ as in Dec-POMDP. On the other hand, since we assume that $\mathcal{T}^*_{VD}$ converges in Dec-POMDP, which means $\prod\nolimits_{VD}$ can find the optimal $f^*_{Dec}(Q_1(s, a_1), \cdots, Q_n(s, a_n))$ to approximate $Q^*$ within the function class $\mathcal{Q}_{tot}$. In Dec-POMDP with VSP, to approximate the same $Q^*$, $\prod\nolimits_{VD}$ simply needs to set $f^*_{VSP} = f^*_{Dec}(Q_1(s, a_1), \cdots, Q_n(s, a_n)) + 0 \times (Q_{1'_c}(s, a_{1'_c}) + \cdots + Q_{n'_c}(s, a_{n'_c}))$. In summary, if the convergence of $\mathcal{T}^*_{VD}$ is guaranteed in a Dec-POMDP, the same guarantee applies to an Dec-POMDP with VSP. In addition, given the same initial joint policy $\bm{\pi}_0$, both $\mathcal{T}_{Dec}^*$ and $\mathcal{T}_{VSP}^*$ can converge to the same MPE $\bm{\pi}_{VSP}^* = \bm{\pi}_{Dec}^*$.
\end{proof}

\begin{appendix_theorem}
	Given an asynchronous decision-making task, define $\mathcal{Q}_{Add}$ as the function class for the additive VD operator $\mathcal{T}_{Add}^*$, and $\mathcal{Q}_{Mul}$ as the function class for the multiplicative interaction VD operator $\mathcal{T}_{Mul}^*$, then:
	\begin{equation}
		\mathcal{Q}_{Add} \subsetneq \mathcal{Q}_{Mul}  \nonumber.
	\end{equation}
\end{appendix_theorem}

\begin{figure}[h]
	\centering
	\includegraphics[width=0.14\textwidth]{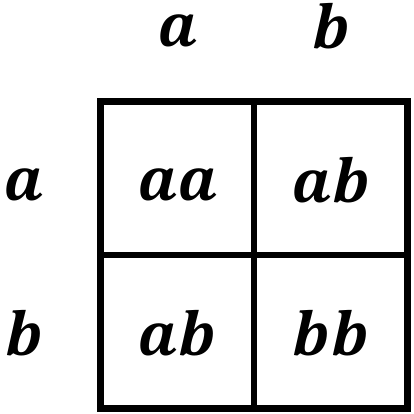}
	\captionof{figure}{Example payoff matrix.}
	\label{fig:asyn_game}
\end{figure}

\begin{proof}
	By comparing Eq. (\ref{eq:ai}) of the additive interactive VD with Eq. (\ref{eq:mvd}) of the multiplicative interactive VD, we clearly see that $\mathcal{Q}_{Add} \subset \mathcal{Q}_{Mul}$. Therefore, the remaining part is to prove the strictness of this inclusion.

	We consider a two-player asynchronous decision-making game where the row agent makes the first move, choosing either action $a$ or $b$, followed by the column agent who can also select action $a$ or $b$. The final reward is determined by multiplying the values of the chosen actions, and the reward matrix is illustrated in Figure \ref{fig:asyn_game}. For multiplicative interactive VD, we set the individual utility as an identity mapping $Q_i(a_i) = a_i$ and let $k_0=k_1=k_2=0$, $k_{12}=1$ to learn the ground truth reward function. However, for additive interactive VD, it is evident that it cannot learn the relationship of multiplying different action values.
\end{proof}

	\section{Derivations}

\subsection{Multiplicative Interaction Form}
\label{app:mi}

According to \cite{qatten}, we can expand $Q_{tot}$ in terms of $Q_i$ as:
\begin{equation}
	\label{app:deduce_tot}
	Q_{tot} = c + \sum_i \mu_i Q_i + \sum_{ij} \mu_{ij} Q_i Q_j + \cdots + \sum_{i_1, \cdots, i_k} \mu_{i_1 \cdots i_k} Q_{i_1} \cdots Q_{i_k} + \cdots,
\end{equation}
where $c$ is a constant and $\mu_{i_1 \cdots i_k} = \frac{1}{k!} \frac{\partial^k Q_{tot}}{\partial Q_{i_1} \cdots \partial Q_{i_k}}$, and then we can expand $Q_i$ near the optimal action $a_i^*$ as:
\begin{equation}
	\label{app:deduce_ind}
	Q_i(a_i) = \alpha_i + \beta_i(a_i - a_i^*)^2 + o((a_i - a_i^*))^2.
\end{equation}

In Dec-POMDP with VSP, we denote the first-order term $\sum_i \mu_i Q_i = \sum_{i_d} \lambda^d_{i_d,1} Q_{i_d} + \sum_{i_c} \lambda^c_{i_c,1} Q_{i_c} + \sum_{i'_c} \lambda^{c'}_{i'_c,1} Q_{i'_c}$. We then apply Eq. (\ref{app:deduce_ind}) into the second-order term in Eq. (\ref{app:deduce_tot}). For brevity, we split $\sum_{i, j} \mu_{ij} Q_i Q_j$ into two parts, $\sum_{i_d, i'_c} \mu_{i_di'_c} Q_{i_d} Q_{i'_c}$ and $\sum_{i_{\bar{d}}, i_{\bar{c}}} \mu_{i_{\bar{d}}i_{\bar{c}}} Q_{i_{\bar{d}}} Q_{i_{\bar{c}}}$, representing the interactions between all pairs of $i_d$ and $i'_c$, and all other interactions in $\{ \mathcal{N}, \mathcal{N}'_c \}$, respectively. For $\sum_{i_{\bar{d}}, i_{\bar{c}}} \mu_{i_{\bar{d}}i_{\bar{c}}} Q_{i_{\bar{d}}} Q_{i_{\bar{c}}}$, let $i_d \in \mathcal{N}_d$, we have:
\begin{equation*}
	\begin{aligned}
		\sum_{i_{\bar{d}}, i_{\bar{c}}} \mu_{i_{\bar{d}}i_{\bar{c}}} Q_{i_{\bar{d}}} Q_{i_{\bar{c}}}
		&= \sum_{i_{\bar{d}}, i_{\bar{c}}} \mu_{i_{\bar{d}}i_{\bar{c}}} (\alpha_{i_{\bar{d}}} + \beta_{i_{\bar{d}}}(a_{i_{\bar{d}}} - a_{i_{\bar{d}}}^*)^2) (\alpha_{i_{\bar{c}}} + \beta_{i_{\bar{c}}}(a_{i_{\bar{c}}} - a_{i_{\bar{c}}}^*)^2) + o(\parallel a - a^* \parallel^2) \\
		&= \sum_{i_{\bar{d}}, i_{\bar{c}}} \mu_{i_{\bar{d}}i_{\bar{c}}} \alpha_{i_{\bar{d}}} \alpha_{i_{\bar{c}}} + \sum_{i_{\bar{d}}, i_{\bar{c}}} \mu_{i_{\bar{d}}i_{\bar{c}}} \alpha_{i_{\bar{c}}} \beta_{i_{\bar{d}}} (a_{i_{\bar{d}}} - a_{i_{\bar{d}}}^*)^2 + \sum_{i_{\bar{d}}, i_{\bar{c}}} \mu_{i_{\bar{d}}i_{\bar{c}}} \alpha_{i_{\bar{d}}} \beta_{i_{\bar{c}}} (a_{i_{\bar{c}}} - a_{i_{\bar{c}}}^*)^2 \\
		&\ \ \ \  + o(\parallel a - a^* \parallel^2) \\
		&= \sum_{i_{\bar{d}}, i_{\bar{c}}} \mu_{i_{\bar{d}}i_{\bar{c}}} \alpha_{i_{\bar{d}}} \alpha_{i_{\bar{c}}} + \sum_{i_{\bar{d}}, i_{\bar{c}}} \mu_{i_{\bar{d}}i_{\bar{c}}} \alpha_{i_{\bar{c}}} (Q_{i_{\bar{d}}} - \alpha_{i_{\bar{d}}}) + \sum_{i_{\bar{d}}, i_{\bar{c}}} \mu_{i_{\bar{d}}i_{\bar{c}}} \alpha_{i_{\bar{d}}} (Q_{i_{\bar{c}}} - \alpha_{i_{\bar{c}}}) \\
		&\ \ \ \ + o(\parallel a - a^* \parallel^2) \\
		&= - \sum_{i_{\bar{d}}, i_{\bar{c}}} \mu_{i_{\bar{d}}i_{\bar{c}}} \alpha_{i_{\bar{d}}} \alpha_{i_{\bar{c}}} + \sum_{i_{\bar{d}}, i_{\bar{c}}} \mu_{i_{\bar{d}}i_{\bar{c}}} \alpha_{i_{\bar{c}}} Q_{i_{\bar{d}}} + \sum_{i_{\bar{d}}, i_{\bar{c}}} \mu_{i_{\bar{d}}i_{\bar{c}}} \alpha_{i_{\bar{d}}} Q_{i_{\bar{c}}} + o(\parallel a - a^* \parallel^2) \\
		&= c_{i_{\bar{d}}i_{\bar{c}}} + \sum_{i \notin \mathcal{N}'_c, i_d} \mu_{ii_d} \alpha_{i} Q_{i_d} + \sum_{i \notin \mathcal{N}_d, i'_c} \mu_{ii'_c} \alpha_{i} Q_{i'_c} + \sum_{i,i_c} \mu_{ii_c} \alpha_{i} Q_{i_c} + o(\parallel a - a^* \parallel^2) \\
		&= c_2 + \sum_{i_d} \lambda^d_{i_d,2} Q_{i_d} + \sum_{i'_c} \lambda^{c'}_{i'_c,2} Q_{i'_c} + \sum_{i_c} \lambda^c_{i_c,2} Q_{i_c} + o(\parallel a - a^* \parallel^2).
	\end{aligned}
\end{equation*}

Therefore, for the second-order term, we get:
\begin{equation*}
	\begin{aligned}
		\sum_{i, j} \mu_{ij} Q_i Q_j
		= c_2 + \sum_{i_d} \lambda^d_{i_d,2} Q_{i_d} + \sum_{i'_c} \lambda^{c'}_{i'_c,2} Q_{i'_c} + \sum_{i_c} \lambda^c_{i_c,2} Q_{i_c} + \sum_{i_d, i'_c} \lambda^{dc'}_{i_di'_c, 2} Q_{i_d} Q_{i'_c} + o(\parallel a - a^* \parallel^2). \\
	\end{aligned}
\end{equation*}

Similarly, we next consider the third-order term $\sum_{i_1, i_2, i_3} \mu_{i_1i_2i_3} Q_{i_1} Q_{i_2} Q_{i_3}$ and obtain:
\begin{equation*}
	\begin{aligned}
		\sum_{i_1, i_2, i_3} \mu_{i_1i_2i_3} Q_{i_1} Q_{i_2} Q_{i_3}
		&= c_3 + \sum_{i_1, i_2 \notin \mathcal{N}'_c, i_d} \mu_{i_1 i_2 i_d} \alpha_{i_1} \alpha_{i_2} Q_{i_d} + \sum_{i_1, i_2 \notin \mathcal{N}_d, i'_c} \mu_{i_1 i_2 i'_c} \alpha_{i_1} \alpha_{i_2} Q_{i'_c} \\
		&\ \ \ \ + \sum_{i_1 \notin \mathcal{N}_d\ \text{or}\ i_2 \notin \mathcal{N}'_c, i_c} \mu_{i_1 i_2 i_c} \alpha_{i_1} \alpha_{i_2} Q_{i_c} + \sum_{i, i_d, i'_c} \mu_{i i_d i'_c} \alpha_{i_d} \alpha_{i'_c} Q_i \\
		&\ \ \ \ + \sum_{i, i_d, i'_c} \mu_{ii_di'_c} \alpha_i Q_{i_d} Q_{i'_c} + o(\parallel a - a^* \parallel^2) \\
		&= c_3 + \sum_{i_d} \lambda^d_{i_d,3} Q_{i_d} + \sum_{i'_c} \lambda^{c'}_{i'_c,3} Q_{i'_c} + \sum_{i_c} \lambda^c_{i_c,3} Q_{i_c} + \sum_{i} \lambda_{dc',3} Q_i \\
		&\ \ \ \ + \sum_{i_d, i'_c} \lambda^{dc'}_{i_di'_c, 3} Q_{i_d} Q_{i'_c} + o(\parallel a - a^* \parallel^2).
	\end{aligned}
\end{equation*}

Consequently, the general form for the $k$-th order term can be expressed as:
\begin{equation*}
	\begin{aligned}
		\sum_{i_1, \cdots, i_k} \mu_{i_1 \cdots i_k} Q_{i_1} \cdots Q_{i_k}
		&= c_k + \sum_{i_d} \lambda^d_{i_d,k} Q_{i_d} + \sum_{i'_c} \lambda^{c'}_{i'_c,k} Q_{i'_c} + \sum_{i_c} \lambda^c_{i_c,k} Q_{i_c} + \sum_{i} \lambda_{dc',k} Q_i \\
		&\ \ \ \ + \sum_{i_d, i'_c} \lambda^{dc'}_{i_di'_c, k} Q_{i_d} Q_{i'_c} + o(\parallel a - a^* \parallel^2). \\
	\end{aligned}
\end{equation*}

We use $\bar{i}_1, \cdots, \bar{i}_k \in \{ \mathcal{N}, \mathcal{N}'_c \}$ to represent that any $\bar{i}_x$ and $\bar{i}_y$ do not simultaneously satisfy $\bar{i}_x \in \mathcal{N}_d$ and $\bar{i}_y \in \mathcal{N}'_c$. So we take:
\begin{equation*}
	\begin{aligned}
		\lambda^d_{i_d,k} &= \sum_{i_1, \cdots, i_{k-1} \notin \mathcal{N}'_c} \mu_{i_1 \cdots i_{k-1} i_d} \alpha_{i_1} \cdots \alpha_{i_{k-1}}, \\
		\lambda^{c'}_{i'_c,k} &= \sum_{i_1, \cdots, i_{k-1} \notin \mathcal{N}_d} \mu_{i_1 \cdots i_{k-1} i'_c} \alpha_{i_1} \cdots \alpha_{i_{k-1}}, \\
		\lambda^c_{i_c,k} &= \sum_{\bar{i}_1, \cdots, \bar{i}_{k-1}} \mu_{\bar{i}_1 \cdots \bar{i}_{k-1} i_c} \alpha_{\bar{i}_1} \cdots \alpha_{\bar{i}_{k-1}}, \\
		\lambda_{dc',k} &= (k-2) \sum_{i_1, \cdots, i_{k-3}, i_d, i'_c} \mu_{i_1 \cdots i_{k-2} i_d i'_c} \alpha_{i_1} \cdots \alpha_{i_{k-3}} \alpha_{i_d} \alpha_{i'_c}, \\
		\lambda^{dc'}_{i_di'_c, k} &= \sum_{i_1, \cdots, i_{k-2}} \mu_{i_1 \cdots i_{k-2} i_d i'_c} \alpha_{i_1} \cdots \alpha_{i_{k-2}}.
	\end{aligned}
\end{equation*}

Therefore, under Dec-POMDP with VSP, we can obtain the multiplicative interaction value decomposition formula Eq. (\ref{eq:mvd}) near the optimal joint action as follows:
\begin{equation}
	\label{app:deduce_final}
	\begin{aligned}
		Q_{tot}
		&\approx \sum_{k} c_k + \sum_{i_d} \sum_{k} (\lambda^d_{i_d,k} + \lambda_{dc',k}) Q_{i_d} + \sum_{i'_c} \sum_{k} (\lambda^{c'}_{i'_c,k} + \lambda_{dc',k}) Q_{i'_c} \\
		&\ \ \ \ + \sum_{i_c} \sum_{k} (\lambda^c_{i_c,k} + \lambda_{dc',k}) Q_{i_c} + \sum_{i_d,i'_c} \sum_{k} \lambda^{dc'}_{i_di'_c, k} Q_{i_d} Q_{i'_c} \\
		&= c + \sum_{i_d} \lambda^d_{i_d} Q_{i_d} + \sum_{i'_c} \lambda^{c'}_{i'_c} Q_{i'_c} + \sum_{i_c} \lambda^c_{i_c} Q_{i_c} + \sum_{i_d, i'_c} \lambda^{dc'}_{i_d, i'_c} Q_{i_d} Q_{i'_c} \\
		&= c + \sum_i^{n+n'_c} \lambda_i Q_i + \sum_{i_d, i'_c} \lambda^{dc'}_{i_d, i'_c} Q_{i_d} Q_{i'_c}.
	\end{aligned}
\end{equation}

The convergence of $\lambda$ just needs mild conditions, like boundedness of $\alpha_i$ or small growth of $\frac{\partial^k Q_{tot}}{\partial Q_{i_1} \cdots \partial Q_{i_k}}$ in terms of $k$.

\subsection{High-Order Interaction Form}
\label{app:hi}

Based on the derivation process of Eq. (\ref{app:deduce_final}), we can further explore the higher-order interaction forms between agents $i_d$ and $i'_c$, and obtain the $K$th-order (where $1 \le K \le n$) interaction value decomposition formula Eq. (\ref{eq:high}) as follows:
\begin{equation*}
	\begin{aligned}
		Q_{tot}
		&\approx \sum_{k} c_k + \sum_{i_d} \sum_{k} (\lambda^{d,K}_{i_d,k} + \lambda^K_{dc',k} + \cdots + \lambda^K_{dc'_1 \cdots c'_{K-1},k}) Q_{i_d} \\
		&\ \ \ \ \ \ \ \ \ \ \ \ \ \ \ + \sum_{i'_c} \sum_{k} (\lambda^{c',K}_{i'_c,k} + \lambda^K_{dc',k} + \cdots + \lambda^K_{dc'_1 \cdots c'_{K-1},k}) Q_{i'_c} \\
		&\ \ \ \ \ \ \ \ \ \ \ \ \ \ \ + \sum_{i_c} \sum_{k} (\lambda^{c,K}_{i_c,k} + \lambda^K_{dc',k} + \cdots + \lambda^K_{dc'_1 \cdots c'_{K-1},k}) Q_{i_c} \\
		&\ \ \ \ \ \ \ \ \ \ \ \ \ \ \ + \sum_{i_d, i'_c} \sum_{k} \lambda^{dc',K}_{i_di'_c, k} Q_{i_d} Q_{i'_c} + \cdots + \sum_{i_d i'_{c,1} \cdots i'_{c,K-1}} \sum_{k} \lambda^{dc'_1 \cdots c'_{K-1},K}_{i_di'_{c,1} \cdots i'_{c,K-1}, k} Q_{i_d} Q_{i'_{c,1}} \cdots Q_{i'_{c,K-1}} \\
		&= c + \sum_{i_d} \lambda^{d,K}_{i_d} Q_{i_d} + \sum_{i'_c} \lambda^{c',K}_{i'_c} Q_{i'_c} + \sum_{i_c} \lambda^{c,K}_{i_c} Q_{i_c} \\
		&\ \ \ \ \ \ \ + \sum_{i_d, i'_c} \lambda^{dc',K}_{i_d, i'_c} Q_{i_d} Q_{i'_c} + \cdots + \sum_{i_d i'_{c,1} \cdots i'_{c,K-1}} \lambda^{dc'_1 \cdots c'_{K-1},K}_{i_di'_{c,1} \cdots i'_{c,K-1}} Q_{i_d} Q_{i'_{c,1}} \cdots Q_{i'_{c,K-1}} \\
		&= c + \sum_i^{n+n'_c} \lambda^K_i Q_i + \sum_{i_d, i'_c} \lambda^{dc',K}_{i_d, i'_c} Q_{i_d} Q_{i'_c} + \cdots + \sum_{i_d i'_{c,1} \cdots i'_{c,K-1}} \lambda^{dc'_1 \cdots c'_{K-1},K}_{i_di'_{c,1} \cdots i'_{c,K-1}} Q_{i_d} Q_{i'_{c,1}} \cdots Q_{i'_{c,K-1}},
	\end{aligned}
\end{equation*}

where:
\begin{equation*}
	\begin{aligned}
		\lambda^{d,K}_{i_d,k} &= \sum_{i_1, \cdots, i_{k-1} \notin \mathcal{N}'_c} \mu_{i_1 \cdots i_{k-1} i_d} \alpha_{i_1} \cdots \alpha_{i_{k-1}}, \\
		\lambda^{c',K}_{i'_c,k} &= \sum_{i_1, \cdots, i_{k-1} \notin \mathcal{N}_d} \mu_{i_1 \cdots i_{k-1} i'_c} \alpha_{i_1} \cdots \alpha_{i_{k-1}}, \\
		\lambda^{c,K}_{i_c,k} &= \sum_{\bar{i}_1, \cdots, \bar{i}_{k-1}} \mu_{\bar{i}_1 \cdots \bar{i}_{k-1} i_c} \alpha_{\bar{i}_1} \cdots \alpha_{\bar{i}_{k-1}}, \\
		\lambda^K_{dc',k} &= (k-2) \sum_{i_1, \cdots, i_{k-3} \notin \mathcal{N}'_c, i_d, i'_c} \mu_{i_1 \cdots i_{k-2} i_d i'_c} \alpha_{i_1} \cdots \alpha_{i_{k-3}} \alpha_{i_d} \alpha_{i'_c}, \\
		\lambda^K_{dc'_1c'_2,k} &= (k-3) \sum_{i_1, \cdots, i_{k-4} \notin \mathcal{N}'_c, i_d, i'_{c,1}, i'_{c,2}} \mu_{i_1 \cdots i_{k-3}i_di'_{c,1}i'_{c,2}} \alpha_{i_1} \cdots \alpha_{i_{k-4}} \alpha_{i_d} \alpha_{i'_{c,1}} \alpha_{i'_{c,2}}, \\
		& \cdots \\
		\lambda^K_{dc'_1 \cdots c'_{K-1},k} &= (k-K) \sum_{i_1, \cdots, i_{k-K-1}, c'_1, \cdots, c'_{K-1}} \mu_{i_1 \cdots i_{k-K}i_di'_{c,1} \cdots i'_{c,K-1}} \alpha_{i_1} \cdots \alpha_{i_{k-K-1}} \alpha_{i_d} \alpha_{i'_{c,1}} \cdots \alpha_{i'_{c,K-1}}, \\
		\lambda^{dc',K}_{i_di'_c, k} &= \sum_{i_1, \cdots, i_{k-2} \notin \mathcal{N}'_c} \mu_{i_1 \cdots i_{k-2} i_d i'_c} \alpha_{i_1} \cdots \alpha_{i_{k-2}}, \\
		\lambda^{dc'_1c'_2,K}_{i_di'_{c,1}i'_{c,2}, k} &= \sum_{i_1, \cdots, i_{k-3} \notin \mathcal{N}'_c} \mu_{i_1 \cdots i_{k-3} i_d i'_{c,1} i'_{c,2}} \alpha_{i_1} \cdots \alpha_{i_{k-3}}, \\
		& \cdots \\
		\lambda^{dc'_1 \cdots c'_{K-1},K}_{i_di'_{c,1} \cdots i'_{c,K-1}, k} &= \sum_{i_1, \cdots, i_{k-K}} \mu_{i_1 \cdots i_{k-K} i_d i'_{c,1} \cdots i'_{c,K-1}} \alpha_{i_1} \cdots \alpha_{i_{k-K}}.
	\end{aligned}
\end{equation*}

\subsection{Practical Implementation Form}
\label{app:pi}

To achieve MVD-based IGM, we only need $Q_{i_d}$ to satisfy IGM. Therefore, we can convert Eq. (\ref{eq:mvd}) into an additive interactive form, $Q_{tot} = b + \bm{W Q_d}$, and then ensure that $\bm{W} > 0$. Specifically:
\begin{equation*}
	\begin{aligned}
		Q_{tot}
		& = k_0 + \sum_i^{n+n'_c} k_i Q_i + \sum_{i_d, i'_c} k_{i_di'_c} Q_{i_d} Q_{i'_c} \\
		& = (k_0 + \sum_{i'_c} k_{i'_c} Q_{i'_c} + \sum_{i_c} k_{i_c} Q_{i_c}) + \sum_{i_d} (k_{i_d} + \sum_{i'_c} k_{i_di'_c} Q_{i'_c}) Q_{i_d}.
	\end{aligned}
\end{equation*}

However, in certain scenarios, as we cannot ensure $Q_{i'_c} > 0$, we need to keep track of the minimum value of $Q_{i'_c}$ during the training process and obtain $Q^{min}_c$:
\begin{equation}
	\label{eq:app_mvd2}
	\begin{aligned}
		Q_{tot}
		& = (k_0 + \sum_{i'_c} k_{i'_c} Q_{i'_c} + \sum_{i_c} k_{i_c} Q_{i_c}) + \sum_{i_d} ((k_{i_d} - \sum_{i'_c} k_{i_di'_c} Q^{min}_c) + \sum_{i'_c} 2 k_{i_di'_c} \frac{Q_{i'_c} + Q^{min}_c}{2}) Q_{i_d}.
	\end{aligned}
\end{equation}

In the specific implementation of MVD, we utilize a hypernetwork $f_i(s)$ to learn the weights in Eq. (\ref{eq:app_mvd2}). Finally, we obtain:
\begin{equation*}
	\begin{aligned}
		Q_{tot}
		& \approx (f_0 + \sum_{i'_c} |f_{i'_c}| Q_{i'_c} + \sum_{i_c} |f_{i_c}| Q_{i_c}) + \sum_{i_d} (|f_{i_d}| + \sum_{i'_c} |f_{i_di'_c}| \frac{Q_{i'_c} + Q^{min}_c}{2}) Q_{i_d},
	\end{aligned}
\end{equation*}
where $|f_{i_d}| + \sum_{i'_c} |f_{i_di'_c}| \frac{Q_{i'_c} + Q^{min}_c}{2} > 0$, so as to achieve MVD-based IGM.

	\section{Pseudo Code}
\label{app:pc}

\begin{algorithm}
	\caption{Multiplicative Value Decomposition}
	\begin{algorithmic}
		\STATE Initialize a set of agents $\mathcal{N} = \{ 1, 2, \cdots, n \}$
		\STATE Initialize networks of individual agents $Q_i(\tau_i, a_i; \theta)$ and target networks $Q'_i(\tau'_i, a'_i; \theta')$
		\STATE Initialize mixing network $Q_{tot}(\widehat{s}, \widehat{\bm{a}}; \vartheta)$ and target network $Q'_{tot}(\widehat{s}, \widehat{\bm{a}}; \vartheta')$
		\STATE Initialize a replay buffer $\mathcal{B}$ for storing episodes
		\STATE Initialize a joint observation $\bm{\tilde{o}}$ and a joint action $\bm{\tilde{a}}$ of all agents

		\REPEAT
		\STATE Initialize a history embedding $h^0_i$ and an action vector $a^0_i$ for each agent
		\STATE Observe each agent’s partial observation $[o_i^1]_{i=1}^n$
		\STATE For each agent $i_c$ executing action, introduce a corresponding virtual proxy $i'_c$
		\STATE Update $\bm{\tilde{o}}_c$, obtain $\widehat{s}^1$ and $[o_i^1]_{i=1}^{n+n'_c}$

		\FOR{ $t = 1 : T$ }
			\STATE Get $\tau_i^t = \{ o_i^t, h_i^{t-1} \}$ for each agent and calculate the individual utility function $Q_i(\tau_i^t, a_i^{t-1})$
			\STATE Get the hidden state $h^t_i$ and select action $a^t_i$ via $Q_i$ with probability $\epsilon$ exploration. For all agents $i'_c$, use an action mask to restrict their choices to $\bm{\tilde{a}}_c$
			\STATE Execute $\widehat{\bm{a}}^t$, update $\bm{\tilde{a}}$
			\STATE Update $\mathcal{N}'_c$ and $\bm{\tilde{o}}_{c}$
			\STATE Receive a reward $\widehat{r}^t$, transition to the next state $\widehat{s}^{t+1}$
		\ENDFOR

		\STATE Store the episode trajectory to $\mathcal{B}$
		\STATE Sample a batch of episodes trajectories with batch size $b$ from $\mathcal{B}$
		
		\FOR{ $t = 1 : T$ }
			\STATE Calculate the global Q-value $Q_{tot}$ via Eq. (\ref{eq:mvd2})
			\STATE Calculate the target $y = \widehat{r}^t + Q'_{tot}$ using target network
		\ENDFOR
		
		\STATE Update $\theta$ and $\vartheta$ by minimizing the loss $\mathcal{L}(\theta, \vartheta) = (Q_{tot} - y)^2$
		\STATE Periodically update $\theta' \leftarrow \theta$, $\vartheta' \leftarrow \vartheta$

		\UNTIL $Q_i(\tau_i, a_i; \theta)$ converges
	\end{algorithmic}  
\end{algorithm}

	\section{Experimental Details}
\label{app:ed}

\subsection{Asynchronous SMAC}

The StarCraft Multi-Agent Challenge (SMAC) \cite{smac} is one of the most widely used benchmarks for evaluating MARL algorithms. At each time step, all agents receive partial observations, perform actions such as moving or attacking, and receive a global reward. To succeed, agents must cooperate effectively to eliminate all enemy units within a predefined time limit. We use the SMAC environment based on unit micromanagement tasks in StarCraft II (version SC2.4.10) and set the built-in AI controlling enemy units to a difficulty level of 7. It is worth noting that results obtained using different game versions may not be directly comparable. Our modified asynchronous variant of SMAC introduces asynchronous actions for allied units. Specifically, we classify allied units into three types, requiring one, two, or three time steps to complete a movement action, respectively. In this paper, we evaluate all algorithms on three representative combat scenarios in SMAC, covering the easy, hard, and super hard difficulty levels. Table \ref{tab:a_sc2} provides a brief introduction to these scenarios and the maximum training steps.

\begin{figure}[h]
	\centering
	\includegraphics[width=0.5\textwidth]{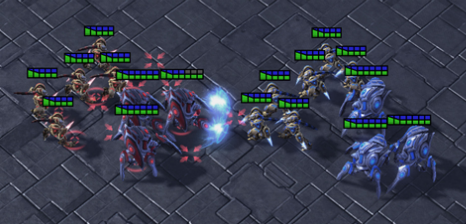}
	\caption{StarCraft Multi-Agent Challenge.}
	\label{fig:sc2_map}
\end{figure}

\begin{table}[h]
	\centering
	\caption{Scenarios Used in asynchronous SMAC Benchmark.}
	\label{tab:a_sc2}
	\begin{tabular}{lllll}
		\toprule
		Map Name & Ally Units & Enemy Units & Total Timesteps & Scenario Type \\
		\midrule
		$2s\_vs\_1sc$  & 2 Stalkers & 1 Spine Crawler & 2M & Easy \\
		$3s5z$  & 3 Stalkers \& 5 Zealots & 3 Stalkers \& 5 Zealots & 2M & Hard \\
		$3s5z\_vs\_3s6z$  & 3 Stalkers \& 5 Zealots & 3 Stalkers \& 6 Zealots & 5M & Super Hard \\
		\bottomrule
	\end{tabular}
\end{table}

\subsection{Overcooked}

In the Overcooked\footnote{The code of Overcooked is from https://github.com/WeihaoTan/gym-macro-overcooked?tab=readme-ov-file.} benchmark, three agents must collaborate to prepare a Tomato-Lettuce-Onion salad and deliver it to the star counter cell as soon as possible. The challenge lies in the fact that the agents are not aware of the correct task sequence beforehand: fetching raw vegetables, placing them on the cut-board cell, chopping them, merging them in a plate, and finally delivering the dish to the star counter. Agents can only observe a limited $5\times5$ grid surrounding themselves. The action space is divided into primitive-action space and macro-action space. Primitive-action space includes five actions that allow agents to move around and complete other tasks: \textit{up}, \textit{down}, \textit{left}, \textit{right}, and \textit{stay}. For example, if an agent holding raw vegetables stops in front of the cut-board cell, it will automatically start chopping. The macro-action space includes ten actions that are combinations of multiple primitive-actions: \textit{Chop}, \textit{Get-Lettuce/Tomato/Onion}, \textit{Get-Plate-1/2}, \textit{Go-Cut-Board-1/2}, \textit{Go-Counter}, \textit{Deliver}. \textit{Chop} refers to cutting a raw vegetable into pieces, which requires the agent to hold vegetables and \textit{stay} in front of the cutting board cell for three time steps. \textit{Get-Lettuce}, \textit{Get-Tomato}, and \textit{Get-Onion} refer to the agent moving to the observed location of raw ingredients and picking up the corresponding vegetables. \textit{Get-Plate-1} and \textit{Get-Plate-2} refer to the agent moving to the observed locations of two plates. \textit{Get-Cut-Board-1} and \textit{Get-Cut-Board-2} refer to the agent moving to the locations of two cut-board cells. \textit{Go-Counter} is only available in map B and refers to the agent moving to the center cell to pick up or put down items. \textit{Deliver} refers to the agent moving to the star counter cell. Chopping vegetables successfully yields a $+10$ reward. Correctly delivering a salad to the star counter cell earns a $+200$ reward, whereas delivering the wrong dish incurs a $-5$ penalty. Furthermore, a $-0.1$ penalty is imposed for each time step. The maximal time steps of an episode is $200$. Each episode terminates if agents successfully deliver a tomato-lettuce-onion salad.

\begin{figure}[h]
	\centering
	\subfloat[Overcooked-A]{
		\includegraphics[width=0.20\textwidth]{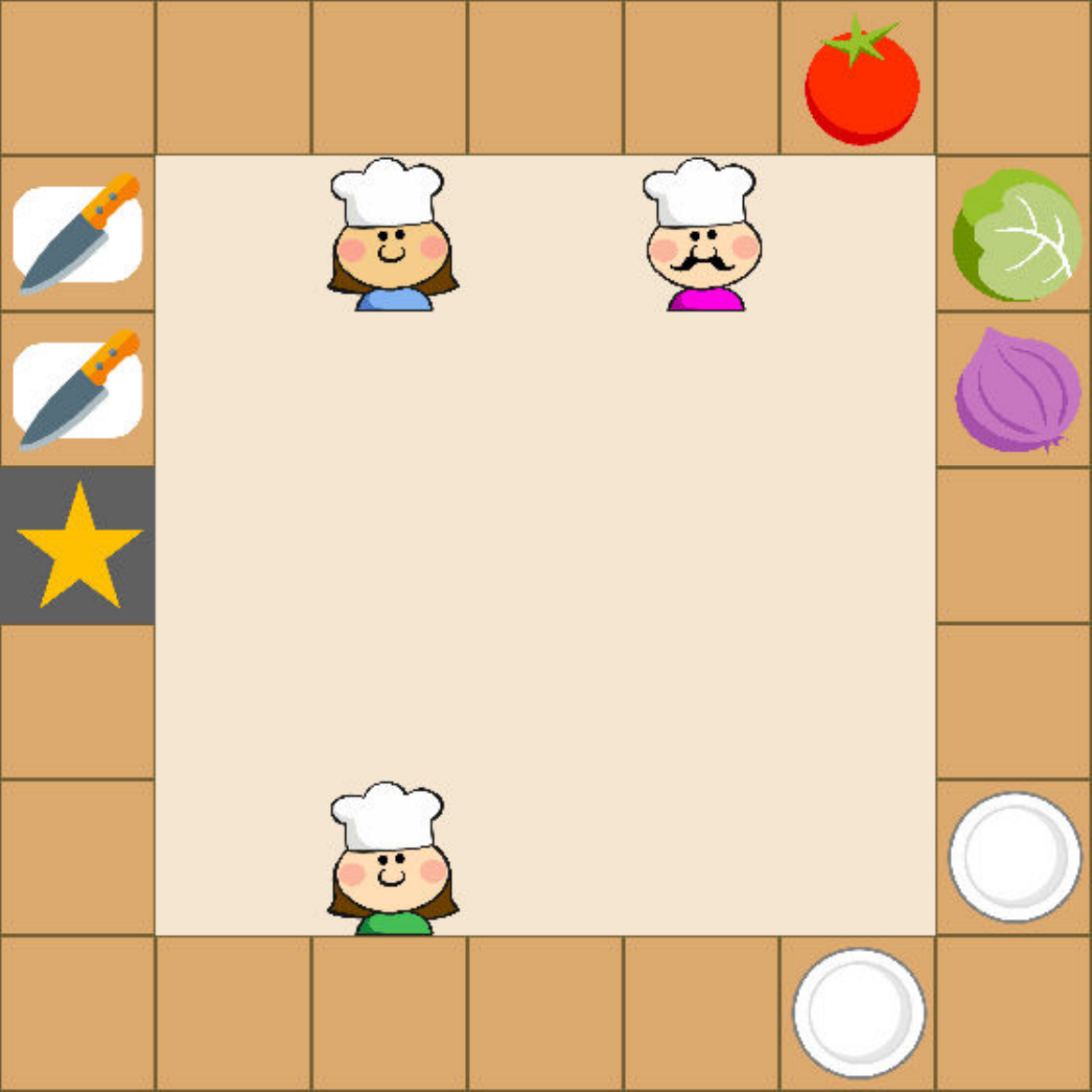}
	}
	\hfil
	\subfloat[Overcooked-B]{
		\includegraphics[width=0.20\textwidth]{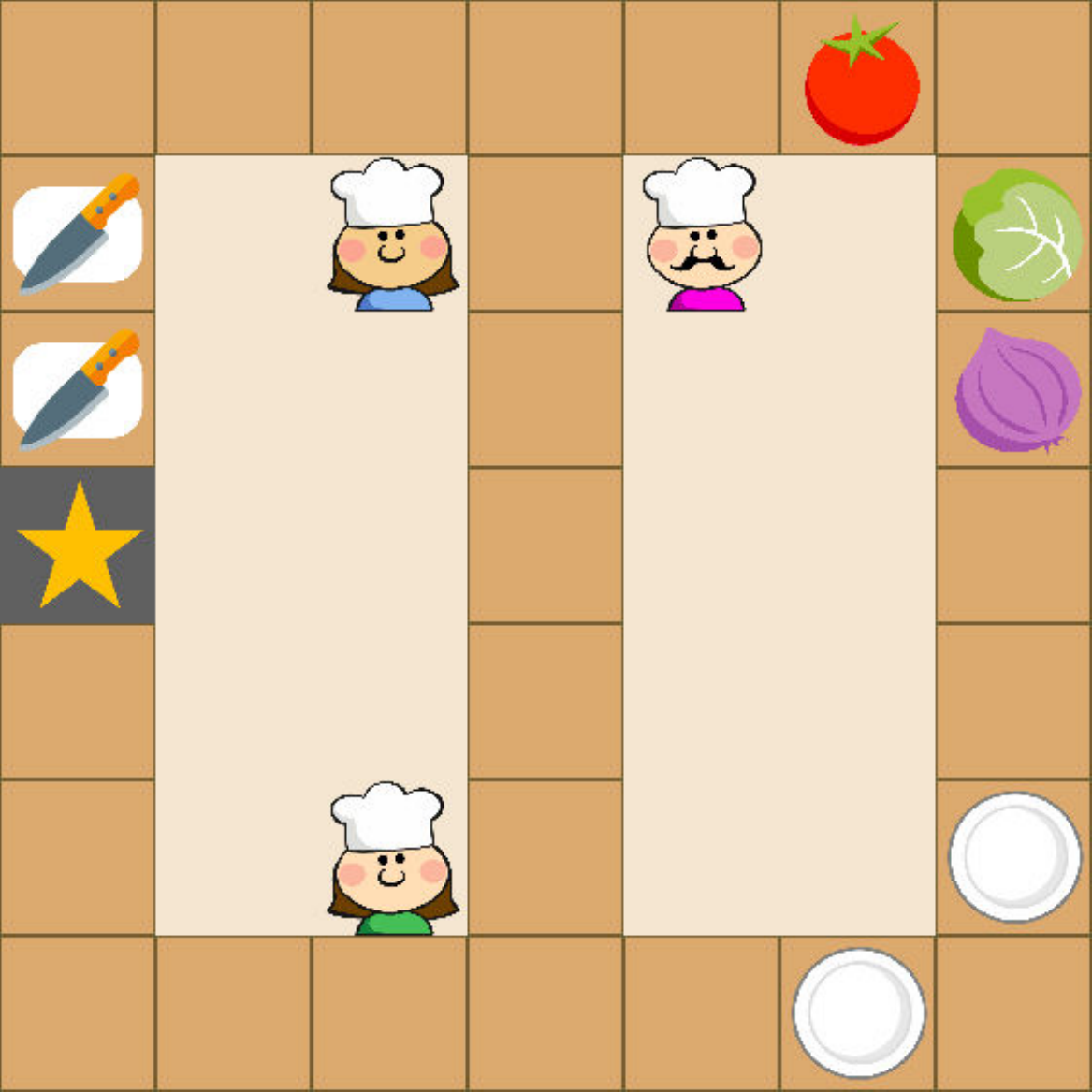}
	}
	\hfil
	\subfloat[Overcooked-C]{
		\includegraphics[width=0.20\textwidth]{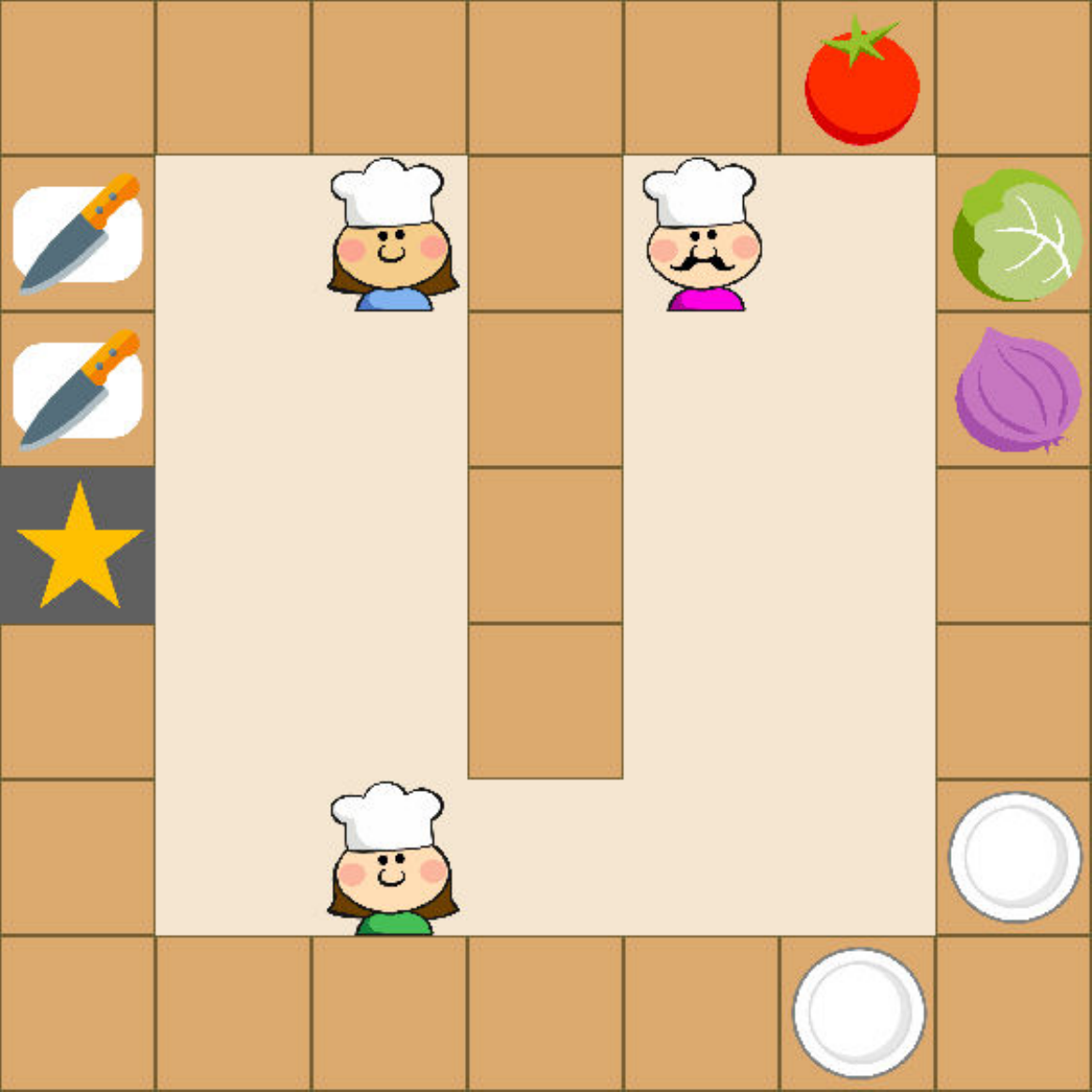}
	}
	\hfil
	\subfloat[Salad Recipe]{
		\includegraphics[width=0.20\textwidth]{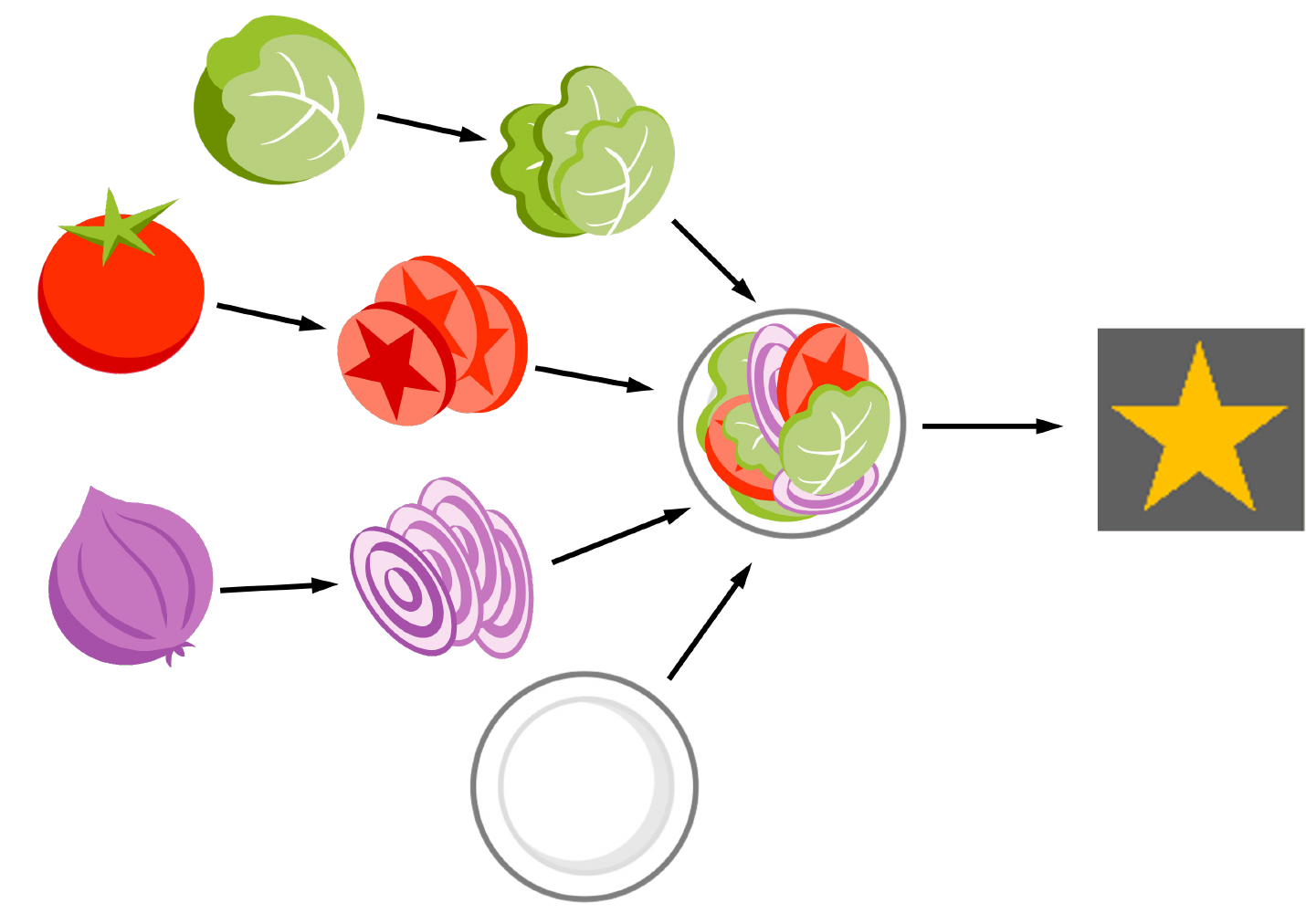}
	}
	\caption{Overcooked environments}
	\label{fig:overcooked_environments}
\end{figure}

\subsection{POAC}

\begin{figure}[h]
	\centering
	\includegraphics[width=0.35\textwidth]{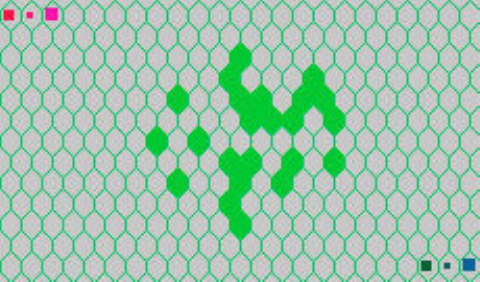}
	\caption{An example of a map, the solid girds are the hidden terrain and the squares are the bots.}
	\label{fig:poac_map}
\end{figure}

Partially observable asynchronous multi-agent cooperation challenge (POAC)\footnote{The code of POAC is from http://turingai.ia.ac.cn/data\_center/show/10.} is a confrontation wargame between two armies, with each army consisting of three heterogeneous agents: infantry, chariot, and tank. These agents possess distinct attributes and different moving speeds, as described in Table \ref{tab:attribution}. The reward is determined by the difference in health loss between allies and enemies. The ultimate objective is to learn asynchronous cooperation strategies to maximize the win rate in every battle, ensuring that agents sustain less damage than their opponents within 600 time steps. All bots are placed on a hexagonal map which contains hidden terrain where it is difficult for enemies to observe when agents are located, as shown in Figure \ref{fig:poac_map}. POAC provides three built-in rule-based bots with different strategies: 1) \textit{KAI0}: an aggressive AI that prioritizes attacking and heads straight for the center of the battlefield. 2) \textit{KAI1}: an AI prefers hiding in special terrain for ambush, a classic wargame tactic. 3) \textit{KAI2}: an AI utilizing guided shooting, where operators squat on special terrain and attack enemies due to the advantage of sight. In this paper, we conduct experiments on the five scenarios consisting of different maps and distinct strategies shown in Table \ref{tab:different_scenarios}. In addition, in the original POAC benchmark, infantry, chariots, and tanks move at speeds of 0.2, 1, and 1, respectively. This means that only infantry movement requires five time steps, with all other actions concluding in one. This limited asynchronicity in the original POAC does not fully demonstrate the strengths of our asynchronous credit assignment method. Therefore, we adjust the movement speeds of infantry, chariots, and tanks to 0.2, 0.3, and 0.5.

\begin{table}[h]
	\centering
	\caption{Operators details.}
	\label{tab:attribution}
	\begin{tabular}{llll}
		\toprule
		Attribution & Infantry & Chariot & Tank \\
		\midrule
		blood  & 7 & 8 & 10 \\
		speed  & 0.2 & 0.3 & 0.5 \\
		observed distance  & 5 & 10 & 10 \\
		attacked distance  & 3 & 7 & 7 \\
		attack damage against tank or chariot  & 0.8 & 1.5 & 1.2 \\
		probability of causing damage against tank or chariot  & 0.7 & 0.7 & 0.8 \\
		attack damage against infantry  & 0.8 & 0.8 & 0.6 \\
		Probability of causing damage against infantry  & 0.6 & 0.6 & 0.6 \\
		shoot cool-down  & 1 & 1 & 1 \\
		shoot preparation time  & 2 & 2 & 0 \\
		can guide shoot  & True & True & False \\
		\bottomrule
	\end{tabular}
\end{table}

\begin{table}[h]
	\centering
	\caption{Operators attributes in different scenarios.}
	\label{tab:different_scenarios}
	\begin{tabular}{llll}
		\toprule
		Scenarios & Map Size & Special Terrain & Guide Shoot \\
		\midrule
		scenario 1  & (13, 23) & False & False \\
		scenario 2  & (13, 23) & True & False \\
		scenario 3  & (17, 27) & True & True \\
		scenario 4  & (27, 37) & True & True \\
		scenario 5  & (27, 37) & True & True \\
		\bottomrule
	\end{tabular}
\end{table}

\subsection{Baselines and Hyperparameters}

We compare our MVD against the following baselines, which can be categorized into three categories:

\begin{enumerate}
	\item The decentralized training and decentralized execution (DTDE) method, IPPO \cite{ippo}, where agents treat other agents as part of the environment, making it suitable for asynchronous tasks. However, in IPPO, each agent is trained independently without considering cooperation among agents and the issue of credit assignment.
	
	\item The discarding type method, MAC IAICC, which is the most advanced algorithm in Xiao et al. (2022). We do not select ASM-PPO \cite{asmppo} and ASM-HPPO \cite{asmhppo} because these two algorithms are only suitable for charging scheduling tasks where each agent has its own reward function, whereas we focus on the credit assignment problem in asynchronous cooperative tasks with shared rewards. Furthermore, we do not select the asynchronous credit assignment algorithm, CAAC \cite{caac}, because its code is not open-source and its application is limited to bus holding control.
	
	\item Five popular credit assignment baselines based on Dec-POMDP with \textbf{blank} padding action, including QMIX \cite{qmix}, Qatten \cite{qatten}, SHAQ \cite{shaq}, ICES \cite{ices}, and NA$^2$Q \cite{na2q} that considers 2nd-order interactions. Additionally, we do not select VarLenMARL \cite{varlenmarl} because this algorithm is also specifically suitable for charging scheduling tasks where each agent has its own reward function.
\end{enumerate}

We implement these baselines and our MVD via PyMARL2\footnote{The code of PyMARL2 is from https://github.com/hijkzzz/pymarl2.}. All hyperparameters follow the code provided by the POAC benchmark and are maintained at a learning rate of 0.0005 by the Adam optimizer, as shown in Table \ref{tab:hyperparameters}. Additionally, for the AC-based methods, we utilize 4 parallel environments for data collection and set the buffer size and batch size to 16. Our experiments are performed on an NVIDIA GeForce RTX 4090 GPU and an Intel Xeon Silver 4314 CPU.

\begin{table}[h]
	\centering
	\caption{Experimental settings of Overcooked and POAC.}
	\label{tab:hyperparameters}
	\begin{tabular}{lll}
		\toprule
		Hyper-parameters & Value & Description \\
		\midrule
		batch size & 32 & number of episodes per update \\
		test interval & 2000 & frequency of evaluating performance \\
		test episodes & 20 & number of episodes to test \\
		buffer size & 5000 & maximum number of episodes stored in memory \\
		discount factor $\gamma$ & 0.99 & degree of impact of future rewards \\
		total timesteps & 5050000 & number of training steps \\
		start $\varepsilon$ & 1.0 & the start $\varepsilon$ value to explore \\
		finish $\varepsilon$ & 0.05 & the finish $\varepsilon$ value to explore \\
		anneal steps for $\varepsilon$ & 50000 & number of steps of linear annealing \\
		target update interval & 200 & the target network update cycle \\
		\bottomrule
	\end{tabular}
\end{table}

In our implementation of MVD, we opted to use the original state $s$ instead of the state $\widehat{s}$ in Dec-POMDP with VSP, as we found that using $\widehat{s}$ did not significantly improve the performance. We employed blank actions as padding actions for Dec-POMDP with VSP. To ensure consistent joint action dimensions in Dec-POMDP with VSP, we always introduced $n$ agents at each time step and utilized an action mask to filter out the invalid agent actions.

	\section{Additional Benchmark Results}
\label{app:all}

\subsection{Asynchronous SMAC}
\label{app:all_sc2}

\begin{figure}[h]
	\centering
	\includegraphics[width=1\textwidth]{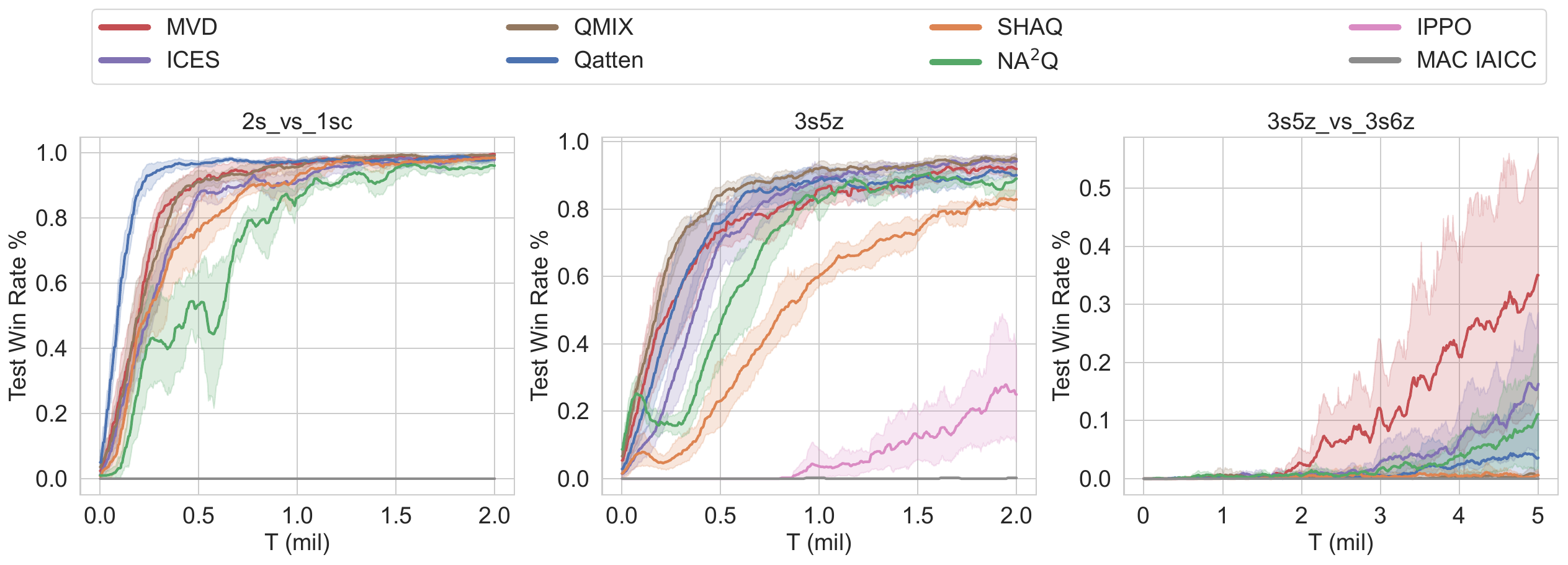}
	\caption{Test win rate \% on three scenarios of asynchronous SMAC benchmark.}
	\label{fig:sc2_all}
\end{figure}

We run experiments on $2s\_vs\_1sc$, $3s5z$, and $3s5z\_vs\_3s6z$ scenarios across the easy, hard, and super hard settings of our modified asynchronous SMAC variant. The results show that in simpler scenarios, both MVD and most baselines converge quickly, while in the super hard scenario, only MVD successfully learns complex asynchronous cooperation strategies. Other baselines either fail to converge or have slow convergence rates. IPPO and MAC IAICC fail to converge across all scenarios due to their disregard for asynchronous credit assignment. NA$^2$Q and SHAQ, affected by padding information, struggle to correctly capture the contributions of asynchronous actions, resulting in lower training efficiency even in simpler scenarios.

\subsection{Overcooked}
\label{app:all_oc}

\begin{figure}[h]
	\centering
	\includegraphics[width=1\textwidth]{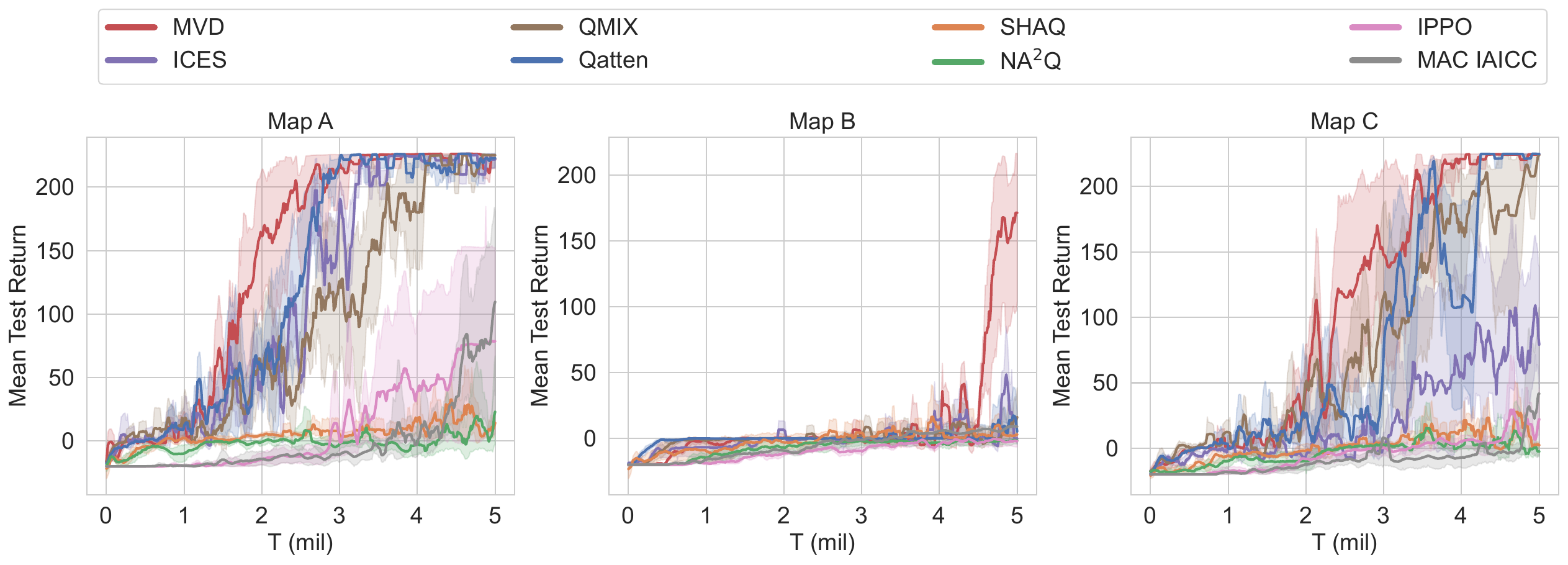}
	\caption{Mean test return on three maps of Overcooked benchmark.}
	\label{fig:oc_all}
\end{figure}

We conducted experiments on all three maps provided by the Overcooked benchmark shown in Figure \ref{fig:overcooked_environments}. The results shown in Figure \ref{fig:oc_all} demonstrate that our MVD outperforms other baselines in different scenarios. IPPO and MAC IAICC, which exhibited slower training speeds on the simple map A, become unable to converge on the more complex and highly collaborative maps B and C. This indicates that discarding decision information from other agents struggles with complex asynchronous cooperation tasks. Due to interference from redundant padding action and complex credit assignment models, NA$^2$Q and SHAQ fail to converge in all scenarios. QMIX and Qatten utilize relatively simple mixing network structures, demonstrating greater robustness to padding action. Consequently, they exhibit comparatively better performance than NA$^2$Q and SHAQ. Map B is divided into two sections, posing special demands on cooperation between agents: Agents in the right section can only prepare raw vegetables and plates, while agents in the left section are restricted to chopping and serving. The cooperation between agents in different sections is weak. This arrangement challenges most baselines to grasp the optimal cooperation strategy.

\subsection{POAC}
\label{app:all_poac}

\begin{figure}[h]
	\centering
	\includegraphics[width=1\textwidth]{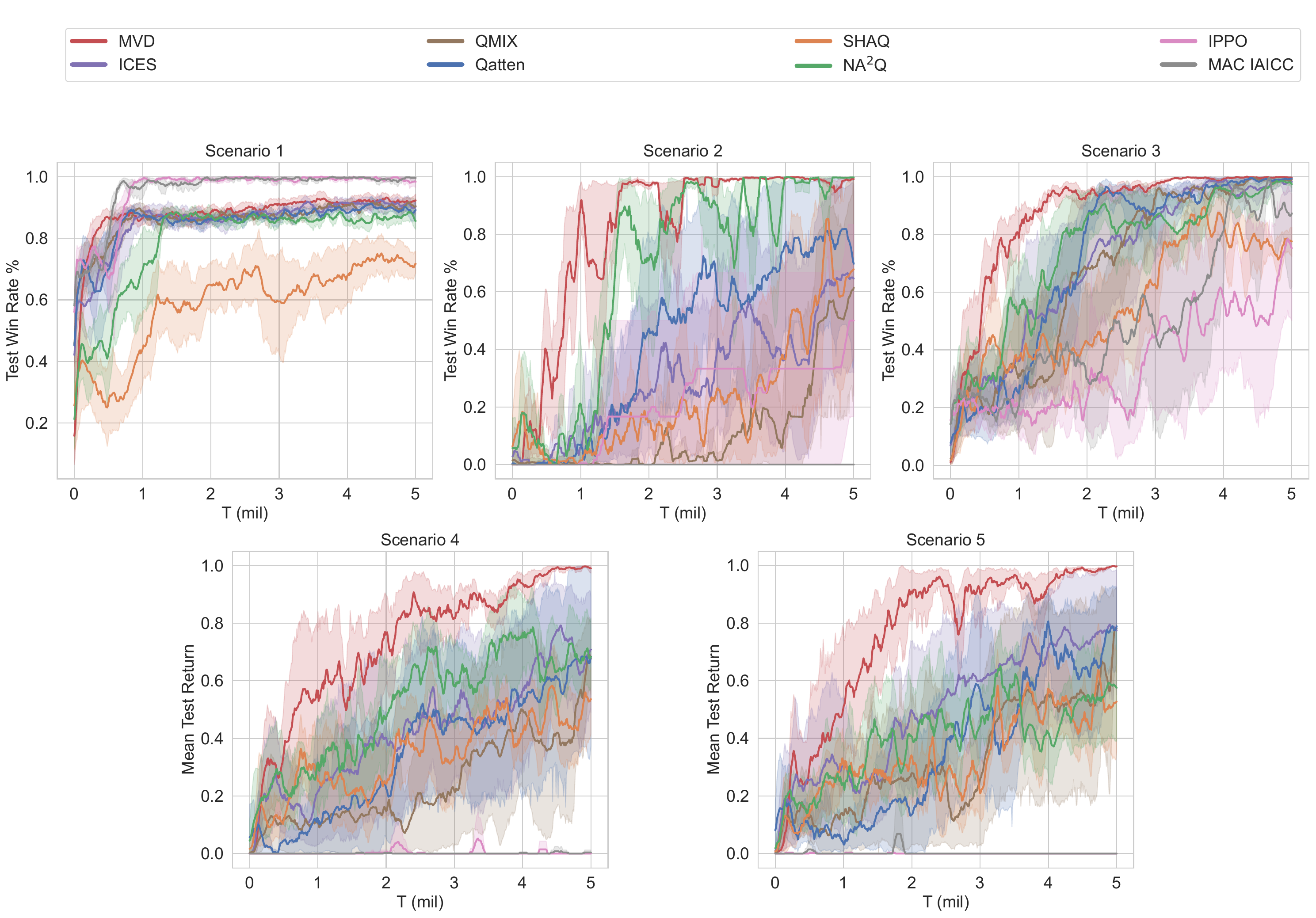}
	\caption{Test win rate \% on all scenarios of POAC benchmark.}
	\label{fig:bq_all}
\end{figure}

As shown in Figure \ref{fig:bq_all}, we compare the performance of our MVD with the baselines in five scenarios provided by the POAC benchmark. It can be observed that MVD remains competitive in simple scenarios 1 and 2. In scenario 1, both IPPO and MAC IAICC achieved optimal performance. This is primarily due to the fact that in this particular setting, a single agent alone is sufficient to defeat the opposing army. Over-considering cooperation among agents may actually lead the algorithm to converge to a sub-optimal strategy. SHAQ requires the explicit computation of the shapley value for each agent. Nonetheless, because padding actions have no actual impact on the environment, this ultimately contributes to its failure. Furthermore, in Scenario 2 where the opposing army adopts a conservative strategy by hiding in special terrains, only the multiplicative interaction in MVD and the GAM in NA$^2$Q provide adequate representational capabilities to locate the enemy's hiding position and learn the optimal joint strategy. Moreover, because MVD employs a simpler model to capture interactions among agents, its training efficiency is higher than that of NA$^2$Q. Conversely, due to limited model representation or exploratory capabilities, the other baselines require more training data to achieve a sophisticated cooperation strategy.

	\section{Additional Ablation Studies}

\subsection{Dec-POMDP with VSP}
\label{app:ablation_adex}

To comprehensively evaluate the advantages of modeling asynchronous decision-making scenarios using Dec-POMDP with VSP, we extend the additive interaction VD algorithms, including QMIX, Qatten, and the high-order interaction VD algorithm NA$^2$Q, to Dec-POMDP with VSP. This allows synchronous credit assignment methods to uniformly consider the marginal contributions of asynchronous decisions from different time steps. We compare the impact of introducing virtual proxies on these VD algorithms.


\begin{figure}[!h]
	\centering
	\includegraphics[width=1\textwidth]{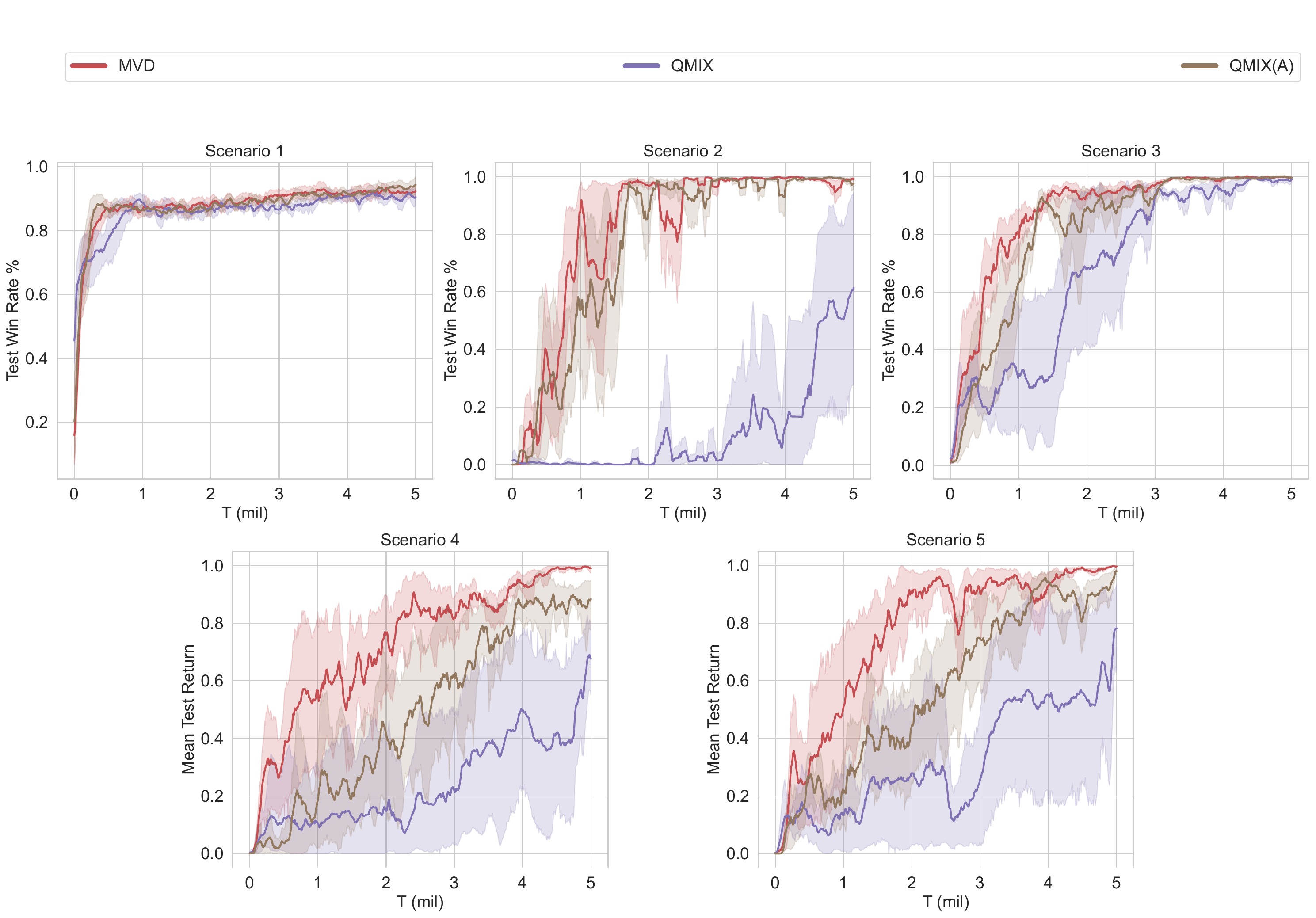}
	\caption{Influence of introducing virtual proxies on QMIX.}
	\label{fig:bq_all_adex_qmix}
\end{figure}

\begin{figure}[!h]
	\centering
	\includegraphics[width=1\textwidth]{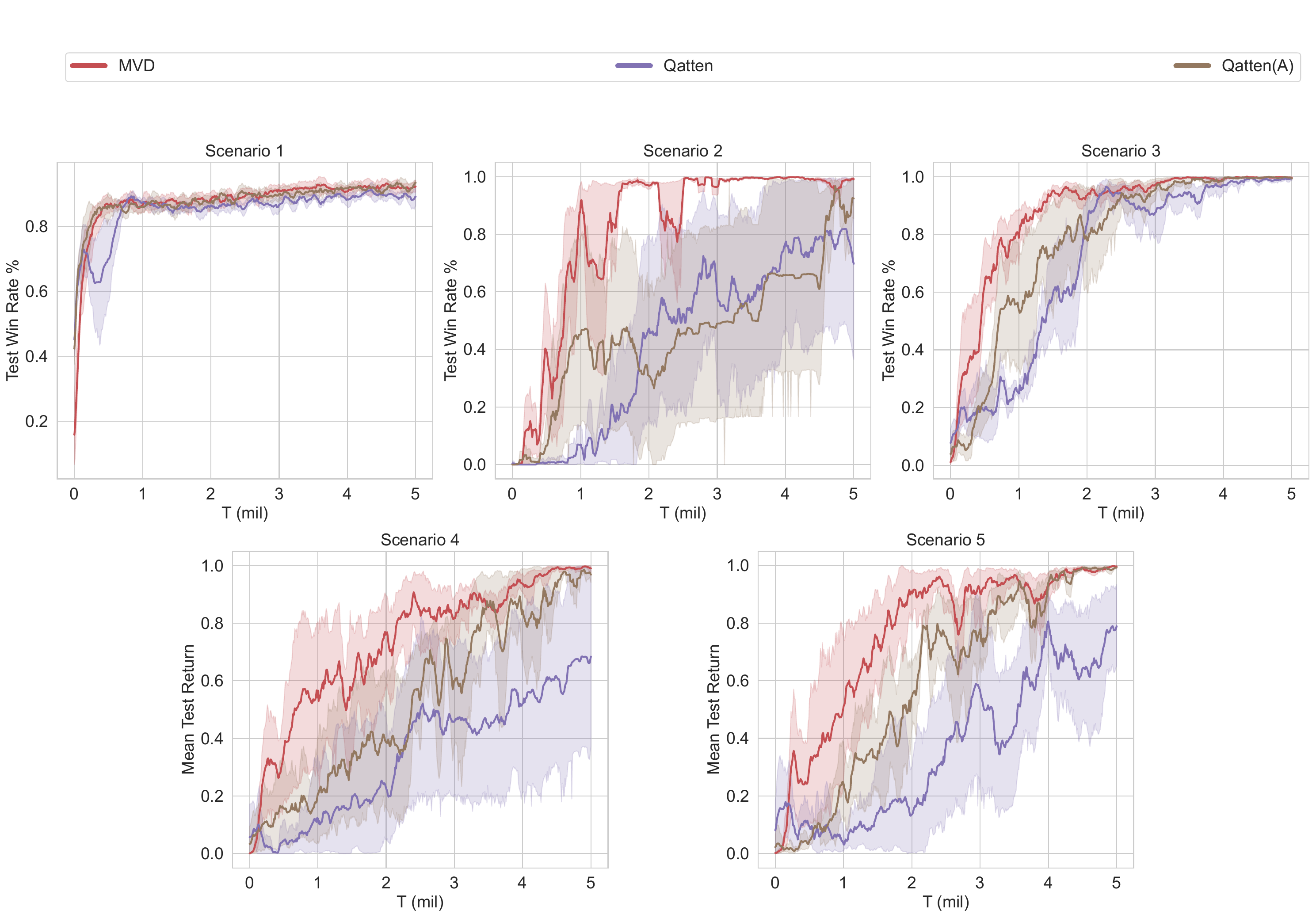}
	\caption{Influence of introducing virtual proxies on Qatten.}
	\label{fig:bq_all_adex_qatten}
\end{figure}

\begin{figure}[!h]
	\centering
	\includegraphics[width=1\textwidth]{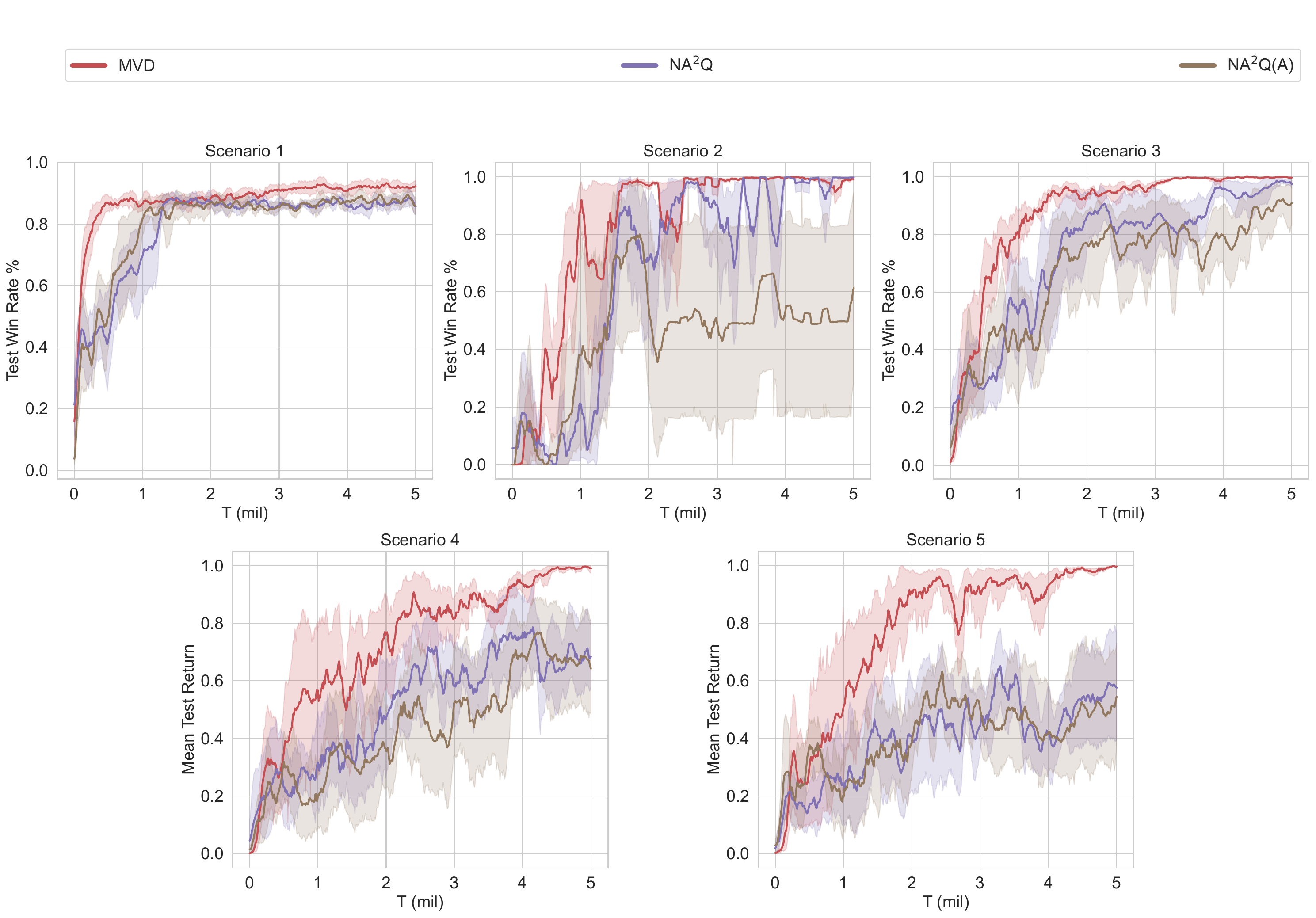}
	\caption{Influence of introducing virtual proxies on NA$^2$Q.}
	\label{fig:bq_all_adex_qnam}
\end{figure}

As shown in Figure \ref{fig:bq_all_adex_qmix} and \ref{fig:bq_all_adex_qatten}, for the additive interaction VD algorithms, QMIX and Qatten, the introduction of virtual proxies significantly improves method performance, especially for QMIX in scenario 2. This suggests that modeling asynchronous decision-making scenarios as Dec-POMDP with VSP facilitates the solution of asynchronous credit assignment problems and enhances training efficiency. However, as shown in Figure \ref{fig:bq_all_adex_qnam}, this is not the case for NA$^2$Q. Introducing virtual proxies can actually bring redundant agent interactions to NA$^2$Q, such as those between agent $i_c$ and agent $i'_c$. This high-order interaction information deteriorates the performance of NA$^2$Q. This implies that for Dec-POMDP with VSP, we must carefully handle the high-order interactions between agents.

\subsection{High-Order Interaction}
\label{app:ablation_high}

\begin{figure}[!h]
	\centering
	\includegraphics[width=1\textwidth]{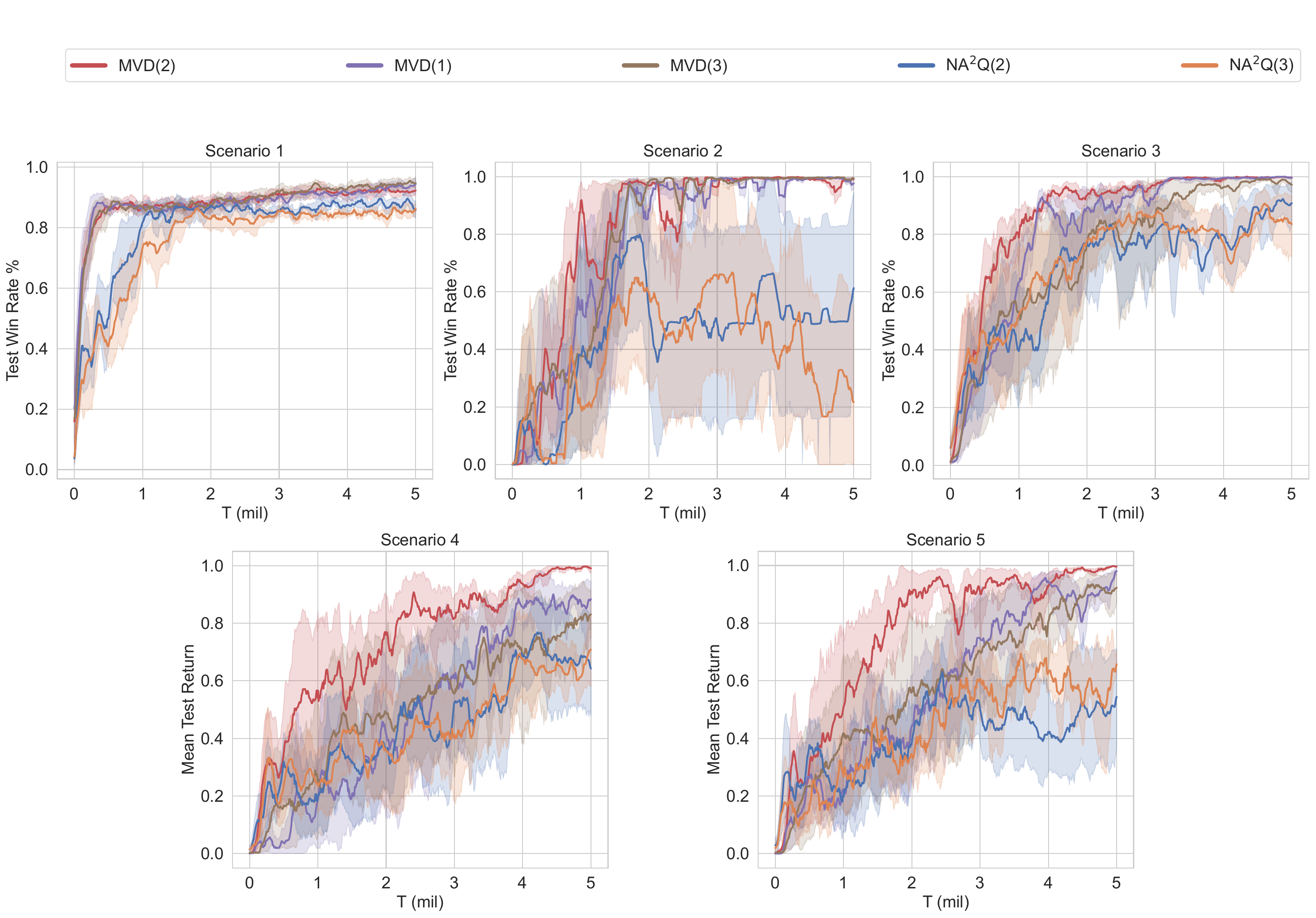}
	\caption{Influence of different order interactions on MVD and NA$^2$Q.}
	\label{fig:bq_all_high}
\end{figure}

To evaluate the impact of higher-order interactions on the performance of VD algorithms in asynchronous settings, we extend NA$^2$Q to Dec-POMDP with VSP and compared MVD and NA$^2$Q under different orders of interaction across five scenarios provided by the POAC benchmark. As shown in Figure \ref{fig:bq_all_high}, MVD(2) which introduces multiplicative interaction to capture the effects of asynchronous decision-making, consistently outperforms MVD(1) which only uses additive interaction. Based on the derivation in Appendix C.2, high-order interactions not only provide deep information on the interplay between agents but also reduce the number of Taylor expansions used in the derivation process, thereby enhancing the accuracy of the VD formula. However, the introduction of high-order interactions complicates the model, thus the performance of MVD(3) remains inferior to MVD(2). In the case of NA$^2$Q, higher-order interactions may slightly improve performance, but they also carry the risk of degrading it. Therefore, in Dec-POMDP with VSP, the interaction between $Q_{i_d}$ and $Q_{i'_c}$ is sufficient to efficiently solve the asynchronous credit assignment problem.

\subsection{Implementation Forms}
\label{app:ablation_f}

\begin{figure}[!h]
	\centering
	\includegraphics[width=1\textwidth]{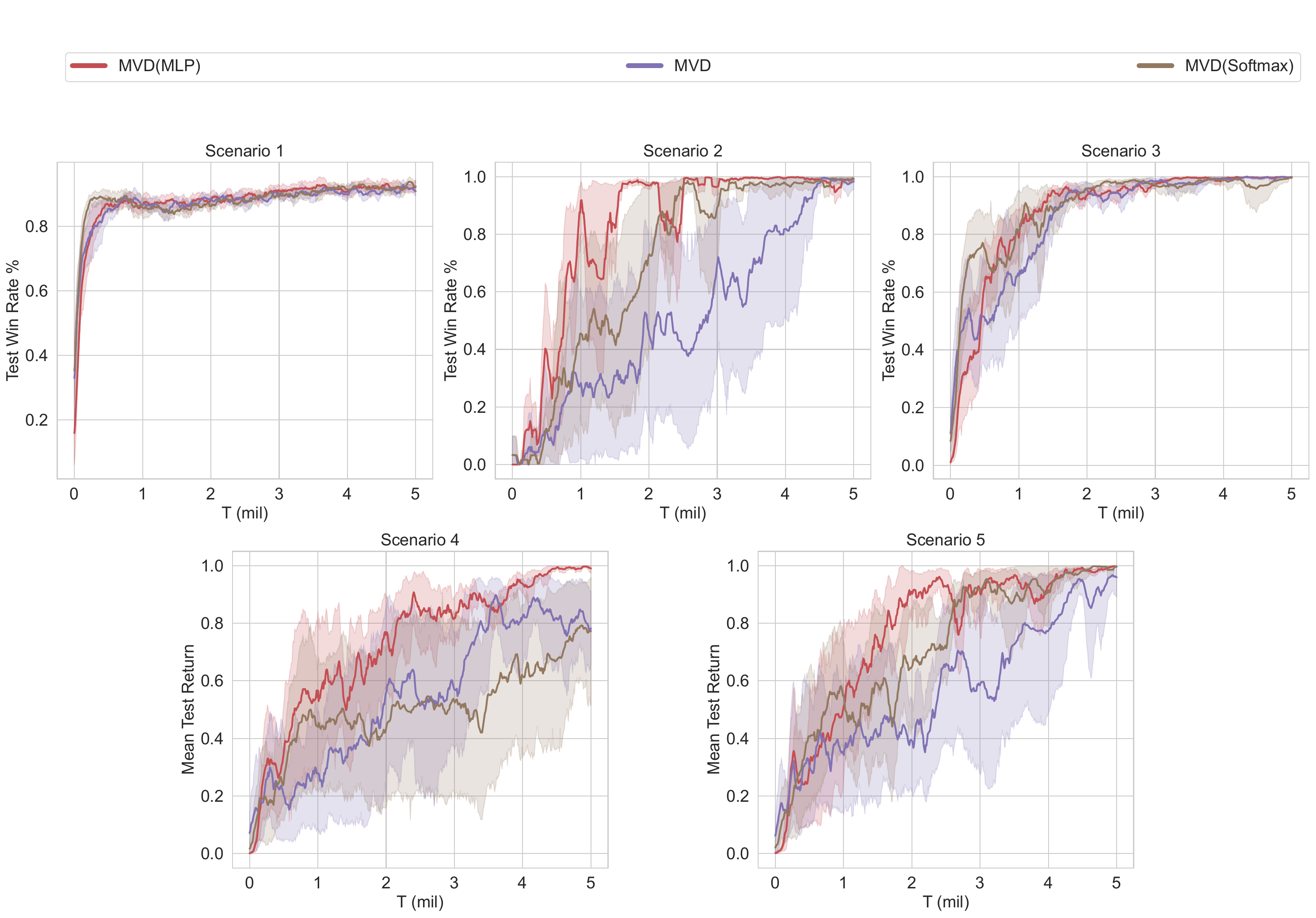}
	\caption{Influence of different implementations on MVD.}
	\label{fig:bq_all_f}
\end{figure}

We compared three different practical implementations of MVD across five scenarios in POAC. As shown in Figure \ref{fig:bq_all_f}, the method of directly obtaining the global Q-value using (\ref{eq:mvd2}) exhibits slower training speed and unstable training process. Although the use of Softmax to implement a multi-head structure can somewhat improve training speed and stability, its performance remains poor in certain scenarios. Therefore, this suggests that for MVD that incorporates multiplicative interactions, the structure of the mixing network should not be overly complex.

\end{document}